\documentclass[vecphys]{svmult}

\usepackage{makeidx}     
\usepackage{graphicx}    
\usepackage[bottom]{footmisc}
\usepackage{latexsym,amsfonts,amsmath,amssymb,graphicx}
\usepackage{epsfig}
\usepackage{colortbl}  
\definecolor{myGray}{gray}{0.9}

\begin{document}

\title*{Elastic Maps and Nets \\ for Approximating Principal Manifolds
\\ and Their Application \\ to Microarray Data Visualization}
\titlerunning{Elastic Maps and Nets}
 \toctitle{Elastic Maps and Nets for Approximating  \ \ \ \ \ \ \ \ \ \ \ \ \ \ \ \ \ \ \ \ \ \ \ \ \  \  \ Principal Manifolds
and Their Application \ \ \ \ \ \ \ \ \ \ \ \ \ \ \ \ \ \ \ \ \ \
\ \ \ \ \ \ \ \ \ \ \ \ \ \ \ \  to Microarray Data Visualization}

\author{Alexander N. Gorban\inst{1,3}\and
Andrei Y. Zinovyev\inst{2,3}}
\authorrunning{A.N. Gorban and A.Y. Zinovyev}
\institute{University of Leicester, University Road, Leicester,
LE1 7RH,  UK, \\ \texttt{ag153@le.ac.uk} \and Institut Curie, 26,
rue d'Ulm, Paris, 75248, France, \\
\texttt{andrei.zinovyev@curie.fr} \and Institute of Computational
Modeling of Siberian Branch of Russian Academy of Sciences,
Krasnoyarsk, Russia}
%
%
\maketitle

\begin{abstract}

Principal manifolds are defined as lines or surfaces passing
through ``the middle'' of data distribution. Linear principal
manifolds (Principal Components Analysis) are routinely used for
dimension reduction, noise filtering and data visualization.
Recently, methods for constructing non-linear principal manifolds
were proposed, including our {\it elastic maps} approach which is
based on a physical analogy with elastic membranes. We have
developed a general geometric framework for constructing
``principal objects'' of various dimensions and topologies with
the simplest quadratic form of the smoothness penalty which allows
very effective parallel implementations. Our approach is
implemented in three programming languages (C++, Java and Delphi)
with two graphical user interfaces (VidaExpert and ViMiDa
applications). In this paper we overview the method of elastic
maps and present in detail one of its major applications: the
visualization of microarray data in bioinformatics. We show that
the method of elastic maps outperforms linear PCA in terms of data
approximation, representation of between-point distance structure,
preservation of local point neighborhood and representing point
classes in low-dimensional spaces.


\end{abstract}

\section{Introduction and Overview}

\subsection{Fr\'echet Mean and Principal Objects: \\ K-Means, PCA, what else?}

A fruitful development of the statistical theory in the 20th
century was an observation made by Fr\'echet in 1948
\cite{Frechet48} about the non-applicability of the mean value
calculation in many important applications when the space of
multi-dimensional measurements is not linear. The simple formula
$M(x)= (1/N)\sum_{i=1}^N x_i$ where $x = \{x_i\}$ does not work if
data points $x$ are defined on a Riemannian manifold (then $M(x)$
can simply not be located on it) or the distance metric defined in
the data space is not quadratic.

This led to an important generalization of the mean value notion
as a set which minimizes mean-square distance to the set of data
samples:

\begin{equation}\label{Frechet}
M_F(x) = \arg\, \min_{y \in D} \sum_i (dist(y,x_i))^2\; ,
\end{equation}

\noindent where $D$ is the space of data samples and $dist(x,y)$ is
a distance function between two points in $D$.

The existence and uniqueness of the \index{Fr\'echet
mean}Fr\'echet mean (\ref{Frechet}) is not generally guaranteed,
but in many applications (one of the most important is statistical
analysis in the shape space, see \cite{Kendall89}) it proves to be
constructive. In vector spaces $M_F=M$ only in the case of
Euclidean (more generally, quadratic) metrics, whereas
(\ref{Frechet}) allows to define a mean value with an alternative
choice of metric.

Let us look at Fr\'echet definition from a different angle. We
would like to utilize such spaces $D$ in (\ref{Frechet}) that does
not necessarily coincide with the space of data samples. Let us
consider a space of objects such that one can measure distance
from an object $y$ to a sample $x_i$. Thus, using (\ref{Frechet})
one can find one (or a set of) ``mean'' or ``principal'' object in
this space with respect to the dataset $X$.

Interestingly, simple examples of ``principal objects'' are
equivalent to application of very well-known methods in statistical
data analysis.

For example, let us consider $D$ as a space of $k$-tuples, i.e.
$k$-element sets of vectors in $R^m$ (centroids), where $m$ is the
dimension of the data space, and define distance between $y \in D$
and a point $x_i \in R^m$ as

\begin{equation}
dist(y,x_i) = \min_{k=1..k}\|y_k-x_i\|\; ,
\end{equation}

\noindent where $y_i$ is the $i$th vector in the $k$-tuple $y$.
Then the  Fr\'echet mean corresponds to the optimal positions of
centroids in the $K$-means clustering problem\footnote{Usually by
$K$-means one means an iterative algorithm which allows us to
calculate the locally optimal position of $k$ centroids. To avoid
confusion, we can say that the $K$-means clustering problem
consists in finding a good approximation to the global optimum and
the algorithm is one of the possible ways to get close to it.}.
Detailed discussion of the data approximation (``recovering")
approach with application to clustering is presented in
\cite{4Mi05}.

Another important example is the space $D$ of linear manifolds
embedded in $R^m$. Let us consider a linear manifold $y$ defined
in the parametric form $y(t) = a+bt$, where $t \in
[-\infty;+\infty]$ is a parameter and $a,b \in R^m$. Let us define
the function $dist(y,x)$ as the distance to the orthogonal
projection of $x$ on $y$ (i.e., to the closest point on $y$):

\begin{equation}
dist(y,x_i) = \min_{t \in [-\infty;+\infty]}\|y(t)-x_i\|\; .
\end{equation}

Then, in the case of Euclidean metrics $\|y(t)-x\|^2 = (y(t)-x)^2$
the ``mean'' manifold $y=M_F(x)$ corresponds exactly to the first
principal component of $X$  (and this is exactly the first Pearson
definition \cite{4Pearson1901}). Analogously, the $k$-dimensional
``mean'' linear manifold corresponds to the $k$-dimensional
principal linear manifold.

The initial geometrical definition of the principal component as a
line minimizing mean square distance from the data points has
certain advantages in application. First, highly effective
algorithms for its computation are proposed in the case of data
spaces with very high dimension. Second, this definition is easily
adapted to the case of incomplete data, as we show below.

In a certain sense, $K$-means clustering and linear principal
manifolds are extreme cases of \index{principal object}
``principal objects'': the first one is maximally flexible with no
particular structure, simply a $k$-tuple, while the second one
presents a rigid linear manifold fitted to data. It is known that
linear principal components (including the mean value itself, as a
0-dimensional principal object) are optimal data estimators only
for gaussian data distributions. In practice we are often faced
with non-gaussivities in the data and then linear PCA is optimal
only in the class of {\it linear} estimators. It is possible to
construct a ``principal object'' in-between these two extreme
alternatives ($K$-Means {\it vs} PCA), such that it would be
flexible enough to adapt to data non-gaussivity and still have
some regular structure.

{\it Elastic nets} \index{elastic net} were developed in a series
of papers
\cite{Gorban01,GorbanRossiev99,GapsGRW,GorbanVisPreprint01,GorbanCHAOS01,Gorban03}.
By their construction, they are system of elastic springs embedded
in the data space. This system forms a regular grid such that it
can serve for approximation of some low-dimensional manifold. The
elastic coefficients of this system allow the switch from
completely unstructured $K$-means clustering (zero elasticity) to
the estimators located closely to linear PCA manifolds (very rigid
springs). With some intermediate values of the elasticity
coefficients, this system effectively approximates non-linear
principal manifolds which are briefly reviewed in the next
section.

\subsection{Principal Manifolds}

Principal manifolds \index{principal manifold} were introduced in
PhD thesis of Hastie \cite{Hastie84}  and then in the paper of
Hastie and Stueltze \cite{HastieStuetzle89} as lines or surfaces
passing through ``the middle'' of the data distribution. This
intuitive definition was supported by a mathematical notion of
self-consistency: every point of the principal manifold is a
conditional mean of all points that are projected into this point.
In the case of datasets only one or zero data points are projected
in a typical point of the principal manifold, thus, one has to
introduce smoothers that become an essential part of the principal
manifold construction algorithms.

Since these pioneering works, many modifications and alternative
definitions of principal manifolds have appeared in the literature.
Theoretically, existence of self-consistent principal manifolds is
not guaranteed for arbitrary probability distributions. Many
alternative definitions were introduced (see, for example,
\cite{KeglThesis99}) in  order to improve the situation and to allow
the construction of principal curves (manifolds) for a distribution
of points with several finite first moments. A computationally
effective and robust algorithmic kernel for principal curve
construction, called the Polygonal algorithm, was proposed by
K\'{e}gl et al. \cite{Kegl99}. A variant of this strategy for
constructing principal graphs was also formulated in the context of
the skeletonization of hand-written digits \cite{Kegl02}.
``Kernalised" version of PCA was developed in \cite{4Scholkopf1998}.
A general setting for constructing regularized principal manifolds
was proposed in \cite{4Smola1999,Smola2001}. In a practical example
\cite{Smola2001} of oil flow data it was demonstrated that
non-linear principal manifolds reveal point class structure better
than the linear ones. Piece-wise construction of principal curves by
fitting unconnected line segments was proposed in \cite{Verbeek00}.

Probably, most scientific and industrial applications of principal
manifold methodology were implemented using the Kohonen
Self-Organizing Maps (SOM) approach developed in the theory of
neural networks \cite{Kohonen82}. These applications are too
numerous to be mentioned here (see reviews in
\cite{SOMbiblioPart1,SOMbiblioPart2,Yin2002,Yin2003}): we only
mention that SOMs and its numerous improvements, indeed, can provide
principal manifold approximations (for example, see
\cite{Mulier95,Ritter92}) and are computationally effective. The
disadvantage of this approach is that it is entirely based on
heuristics; also it was shown that in the SOM strategy there does
not exist any objective function that is minimized by the training
process \cite{Erwin92}. These limitations were discussed in
\cite{4Bishop1998}, and the Generative Topographic Mapping (GTM)
approach was proposed, in which a Gaussian mixture model defined on
a regular grid and the explicit log likelihood cost function were
utilized.

Since 1998 we have developed an approach for constructing
principal manifolds based on a physical analogy with elastic
membranes modeled by a system of springs. The idea of using the
elastic energy functional for principal manifold construction in
the context of neural network methodology was proposed in the mid
1990s (see \cite{Gorban98,GorbanRossiev99} and the bibliography
there). Due to the simple quadratic form of the curvature penalty
function, the algorithm proposed for minimization of elastic
energy proved to be very computationally effective and allowed
parallel implementations. Packages in three programming languages
(C++, Java and Delphi) were developed
\cite{Elmap,VidaExpert,VIMIDA} together with graphical user
interfaces (VidaExpert \cite{VidaExpert} and ViMiDa
\cite{VIMIDA}). The elastic map approach led to many practical
applications, in particular in data visualization and missing data
value recovery. It was applied for visualization of economic and
sociological tables
\cite{GorbanVisPreprint01,GorbanCHAOS01,GorbanInfo00,ZinovyevBook00},
to visualization of natural \cite{ZinovyevBook00} and genetic
texts \cite{Gorban03,GorbanOpSys03,Zinovyev02}, recovering missing
values in geophysical time series \cite{GapsDGRKM}. Modifications
of the algorithm and various adaptive optimization strategies were
proposed for modeling molecular surfaces and contour extraction in
images \cite{GorbanZinovyev05Computing}. Recently the elastic maps
approach was extended for construction of cubic complexes (objects
more general than manifolds) in the framework of topological
grammars \cite{Gorban07AML}. The simplest type of grammar (``add a
node'', ``bisect an edge'') leads to construction of branching
principal components (principal trees) that are described in the
accompanying paper.

\subsection{Elastic Functional and Elastic Nets}

Let $G$ be a simple undirected graph with set of vertices $Y$ and
set of edges $E$. For $k\geq 2$ a $k$-star in $G$ is a subgraph
with $k+1$ vertices $y_{0,1,...,k} \in Y$ and $k$ edges
$\{(y_0,y_i) | i=1,..,k \} \subset E$. Suppose that for each $k
\geq 2$, a family $S_k$ of $k$-stars in $G$ has been selected.
Then we define an {\it elastic graph} as a graph with selected
families of $k$-starts $S_k$ and for which for all $E^{i} \in E$
and $S^{j}_k \in S_k$, the corresponding elasticity moduli
$\lambda_i
> 0$ and $\mu_{kj}> 0$ are defined.

Let $E^{(i)}(0)$, $E^{(i)}(1)$ denote two vertices of the graph
edge $E^{(i)}$ and $S_k^{(j)}(0),...,S_k^{(j)}(k)$ denote vertices
of a $k$-star $S_k^{(j)}$ (where $S_k^{(j)}(0)$ is the central
vertex, to which all other vertices are connected). Let us
consider a map $\phi : Y \rightarrow R^m$ which describes an
embedding of the graph into a multidimensional space. The
\index{elastic energy} {\it elastic energy of the graph embedding}
is defined as

\begin{eqnarray}\label{ElasticFunctional}
U^{\phi}(G) := U^{\phi}_E(G) + U^{\phi}_R(G)\; ,
\\
U^{\phi}_E(G) := \sum_{E^(i)}\lambda_i \left\|
\phi(E^{(i)}(0))-\phi(E^{(i)}(1)) \right\|^2\; ,
\\
 U^{\phi}_R(G) :=
\sum_{S^{(j)}_k} \mu_{kj}\left\| \sum_{i=1}^k\phi(S_k^{(j)}(i))-k
\phi(S_k^{(j)}(0)) \right\|^2\; .
\end{eqnarray}

\begin{figure}[t]
\centering{
\includegraphics[width=100mm, height=50mm]{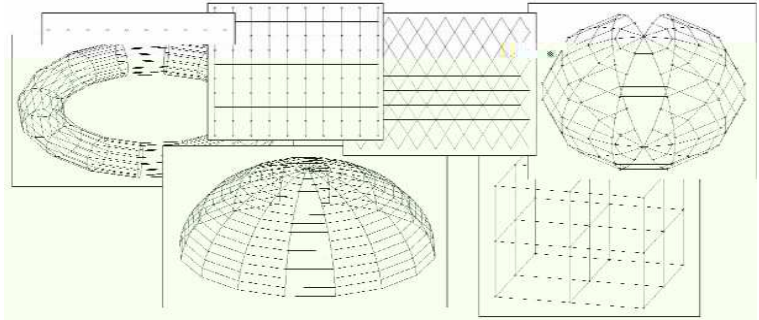}
} \caption{Elastic nets used in practice \label{4Fig2}}
\end{figure}

This general definition of an \index{elastic graph} elastic graph
and its energy is used in \cite{Gorban07AML} and in the
accompanying paper to define elastic cubic complexes, a new
interesting type of principal objects. In this paper we use
elastic graphs to construct \index{grid approximation}grid
approximations to some regular manifolds. Because of this we limit
ourselves to the case $k=2$ and consider graphs with nodes
connected into some regular grids (see Fig.~\ref{4Fig2}). In this
case we will refer to $G$ as an {\it elastic net}\footnote{To
avoid confusion, one should notice that the term elastic net was
independently introduced by several groups: in \cite{Durbin87} for
solving the traveling salesman problem, in \cite{GorbanCHAOS01} in
the context of principal manifolds and recently in \cite{Hastie05}
in the context of regularized regression problem. These three
notions have little to do with each other and denote different
things.}.

\begin{figure}[t]
\centering{
\includegraphics[width=75mm, height=8mm]{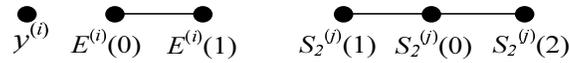}
} \caption{A node, an edge and a rib (2-star) \label{4Fig1}}
\end{figure}

The $2$-star is also called a \index{rib} {\it rib} (see
Fig.\ref{4Fig1}). The contribution \linebreak $\mu_{j}\left\|
\phi(S^{(j)}_2(1))+\phi(S^{(j)}_2(2))-2 \phi(S^{(j)}_2(0))
\right\|^2$ (in which one can recognize a discrete approximation
of the second derivative squared) of the $j$th rib to the elastic
energy equals zero only if the embedding of the central vertice is
the average of its neighbours. Ribs introduce a simple quadratic
\index{penalty}penalty term on the non-linearity (in other words,
complexity) of graph embedding.

How is the set ${S_2}$ chosen for an elastic net? In most of the
cases this choice is natural and is dictated by the structure of the
net. It is useful to consider another embedding of the graph $G$ in
a low-dimensional space, where the graph has its ``natural'' form.
For example, on Fig.~\ref{4Fig2} the regular rectangular grid is
embedded into $R^2$ and $S_2$ is defined as all triples of adjacent
collinear vertices.

We would like to find such a map $\phi$ that has low elastic
energy and simultaneously approximates a set of data points $X$,
where the approximation error is defined as the usual mean-square
distance to the graph vertex positions in the data space. For a
given map $\phi:Y \rightarrow R^m$ we divide the data set
$X=\{x_i\}$ into subsets $\{K^{y_j}\}$, where $y_j \in Y$ is a
vertex of the graph $G$. The set $K^{y_j}$ contains data points
for which the node embedding $\phi(y_j)$ is the closest one:

\begin{equation}\label{taxondef}
K^{y_i} = \{x_i| y_j = \arg \min_{y_k\in Y}\|y_k - x_i\| \} \; .
\end{equation}

\noindent Then the approximation error (``approximation energy")
is defined as

\begin{equation}\label{approxerror}
U^\phi_{A}(G,X) = \frac{1}{\sum_{ x \in X} w(x)}\sum_{y \in Y}
\sum_{ x \in K^y} w(x) \|x- \phi(y)\|^2\; ,
\end{equation}

\noindent where $w(x)\geq 0$ are the point weights. In the
simplest case $w(x)=1$ but it might be useful to make some points
`heavier' or `lighter' in the initial data.

The normalization factor ${1}/{\sum_{ x \in X} w(x)}$ in
(\ref{approxerror}) is needed for the law of large numbers: if $X$
is an i.i.d. sample for a distribution with probability density
$\rho(x)$ and finite first two moments then there should exist a
limit of $U^{\phi}_A(G,X)$ for $|X| \to \infty$. This can be easily
proved for $w(x) \equiv 1$ or if $w(x)\geq 0$ is a bounded
continuous function of point $x$. In the limit $|X| \to \infty$ the
energy of approximation is

\begin{equation}\label{LLNapproxerror}
U^\phi_{A}(G,\rho) =\frac{1}{\mathbf{E}(w)}\int_{ x \in V^y} w(x)
\rho(x) \|x- \phi(y)\|^2 \, \D ^m x\; ,
\end{equation}
where $V^y = \{x \in R^m \, | y= \arg \min_{z \in Y}\|z - x\| \}$
is the \index{Voronoi cell}Voronoi cell and $\mathbf{E}(m)= \int
w(x) \rho(x)\, \D ^m x$ is the expectation.

The optimal map $\phi_{opt}$ minimizes the total energy function
(approximation+elastic energy):

\begin{equation}\label{FullFunctional}
U^\phi := U^\phi_A(G,X)+U(G)^\phi \; .
\end{equation}

\begin{figure}[t]
\centering{
\includegraphics[width=90mm, height=50mm]{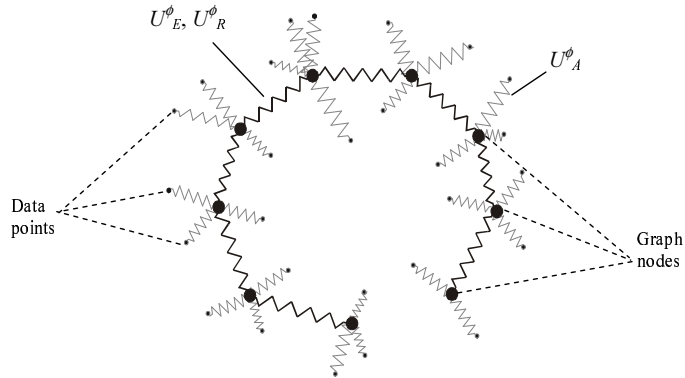}
} \caption{Energy of elastic net \label{4Fig3}}
\end{figure}

To have a visual image of (\ref{FullFunctional}), one can imagine
that every graph node $y_j$ is connected by elastic bonds to its
closest data points in $K^{y_j}$ and simultaneously to the adjacent
nodes (see Fig.~\ref{4Fig3}).

The values $\lambda _{i}$ and $\mu _{j}  $ are the coefficients of
stretching elasticity of every edge $E^{{\rm (}i{\rm )}}$ and of
bending elasticity of every rib $S^{j}_2$. In the simplest case we
have \index{elasticity coefficients}

\[
\lambda _{1} = \lambda _{2} = ... = \lambda _{s} = \lambda (s),
\quad \mu _{1} = \mu _{2} = ... = \mu _{r} = \mu (r)\; ,
\]

\noindent where $s$ and $r$ are the numbers of edges and ribs
correspondingly. To obtain $\lambda (s)$ and $\mu (r)$
dependencies we simplify the task and consider the case of a
regular, evenly stretched and evenly bent grid. Let us consider a
lattice of nodes of ``internal'' dimension $d$ ($d=1$ in the case
of a polyline, $d=2$ in case of a rectangular grid, $d=3$ in the
case of a cubical grid and so on). Let the ``volume'' of the
lattice be equal to $V$. Then the edge length equals $( V /
s)^{1/d}$. Having in mind that for typical regular grids $r
\approx s$, we can calculate the smoothing parts of the
functional: $U_{E}$\textit{$\sim $~}$\lambda s^{{\frac{{d -
2}}{{d}}}}$, $U_{R}$\textit{$\sim $~}$\mu r^{{\frac{{d -
2}}{{d}}}}$. Then in the case where we want $U_R, U_E$ to be
independent on the grid ``resolution'',

\begin{equation}
\label{4eq4} \lambda = \lambda _{0} s^{{\frac{{2 - d}}{{d}}}},
\quad \mu = \mu _{0} r^{{\frac{{2 - d}}{{d}}}}\; ,
\end{equation}

\noindent where $\lambda _{ 0}$, $\mu _{ 0}$ are elasticity
parameters. This calculation is not applicable, of course, for the
general case of any graph. The dimension in this case can not be
easily defined and, in practical applications, the $\lambda _{i}$,
$\mu _{i} $ are often made different in different parts of a graph
according to some adaptation strategy (see below).

The elastic net approximates the cloud of data points and has
regular properties. Minimization of the $U^\phi_A$ term provides
approximation, the $U_E^\phi$ penalizes the total length (or,
indirectly, ``square'', ``volume'', etc.) of the net and
$U_R^\phi$ is a smoothing term, preventing the net from folding
and twisting.

Other forms of $U_R^\phi$ are also used, in particular in the
Polygonal algorithm \cite{Kegl99}. There exists a simple
justification for the quadratic form of $U_R^\phi$ in
(\ref{ElasticFunctional}): it is the simplest non-linearity
penalty (which corresponds to a simple approximation of the second
derivative). All other variants might be more effective in various
specific applications but will be more and more close to our
choice of $U_R^\phi$ with increasing grid resolution.

\section{Optimization of Elastic Nets for Data Approximation}

\subsection{Basic Optimization Algorithm}
\label{subsec:basOptAlg}

Finding the globally optimal map $\phi$ is not an easy problem. In
practice we can apply the splitting optimization strategy similar
to $K$-means clustering. The optimization is done iteratively,
with two steps at each iteration: 1) Determining $\{K^{y_{j}}\}$
for a given $\phi$ and 2) optimizing $\phi$ for a given
$\{K^{y_{j}}\}$.

In order to perform the map optimization step 2 we derive the system
of algebraic linear equations to be solved. Data points are
separated in $K^{y_{j}}, \; j = 1\ldots p$.

Let us denote

\[
\Delta (x,y) = {\left\{ {\begin{array}{l}
 {1,\; x = y} \\
 {0,\; x \ne y\; ,} \\
 \end{array}} \right.}
\]

\[
\Delta E^{ij} \equiv \Delta (E^{(i)}(0),y^{(j)}) - \Delta
(E^{(i)}(1),y^{(j)})\; ,
\]

\[
\Delta R^{ij} \equiv \Delta (S^{(i)}(2),y^{(j)}) + \Delta
(S^{(i)}(1),y^{(j)}) - 2\Delta (S^{(i)}(0),y^{(j)})\; .
\]
That is, $\Delta E^{ij}=1$ if $y^j=E^{(i)}(0)$, $\Delta E^{ij}=-1$
if $y^j=E^{(i)}(0)$, and $\Delta E^{ij}=0$ for all other $y^j$;
$\Delta R^{ij}=1$ if $y^j=R^{(i)}(1)$ or $y^j=R^{(i)}(2)$, $\Delta
R^{ij}=-2$ if $y^j=R^{(i)}(0)$, and $\Delta R^{ij}=0$ for all other
$y^j$. After a short calculation we obtain the system of $p$
\textit{linear} equations to find new positions of nodes in
multidimensional space \{$\phi(y^{i}), i=$1\textit{\ldots p}\}:

\[
{\sum\limits_{k = 1}^{p} a_{jk} \phi(y^{(k)}) = {\frac{{1}}{\sum_{
x \in D} w(x)} }}{\sum\limits_{x \in K^{y_{j}}} ^{} {w(x) x}}\; ,
\]

where

\begin{equation}\label{matrix}
 a_{jk} = \frac{{n_{j} \delta _{jk}}} {\sum_{
x \in D} w(x)} + e_{jk} + r_{jk}\;  , \quad j = 1\ldots p\;  ,
\end{equation}

\noindent $\delta _{jk}$ is Kronecker's $\delta$, and $n_{j}  =
\sum_{x \in K^{y_{j}}} {w(x)}  $, $e_{jk} = {\sum\limits_{i =
1}^{s} {\lambda _{i} \Delta E^{ij}\Delta E^{ik}} }$, $r_{jk} =
{\sum\limits_{i = 1}^{r} {\mu _{i} \Delta R^{ij}\Delta R^{ik}} }$.
The values of $e_{jk}$ and $r_{jk}$ depend only on the structure
of the grid. If the structure does not change then they are
constant. Thus only the diagonal elements of the matrix
(\ref{matrix}) depend on the data set. The $a$ matrix has sparse
structure for a typical grid used in practice. In practice it is
advantageous to calculate directly only non-zero elements of the
matrix and this simple procedure was described in
\cite{GorbanZinovyev05Computing}.

The process of minimization is interrupted if the change in
$U^\phi$ becomes less than a small value $\varepsilon$ or after a
fixed number of iterations.

\begin{figure}[t]
\centering{
\includegraphics[width=80mm, height=25mm]{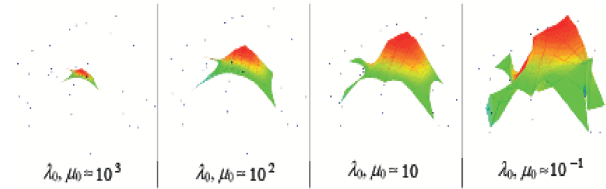}
} \caption{Training elastic net in several epochs (softening)
\label{4Fig4}}
\end{figure}

As usual, we can only guarantee that the algorithm leads to a local
minimum of the functional. Obtaining a solution close to the global
minimum can be a non-trivial task, especially in the case where the
initial position of the grid is very different from the expected (or
unknown) optimal solution. In many practical situations
\index{softening} the ``softening'' strategy can be used to obtain
solutions with low energy levels. This strategy starts with
``rigid'' grids (small length, small bending and large $\lambda $,
$\mu $ coefficients) at the beginning of the learning process and
finishes with soft grids (small $\lambda $, $\mu $ values),
Fig.~\ref{4Fig4}. Thus, the training goes in several epochs, each
epoch with its own grid rigidness. The process of ``softening'' is
one of numerous heuristics  that pretend to find the global minimum
of the energy $U$ or rather a close configuration.

Nevertheless, for some artificial distributions (like the spiral
point distribution, used as a test in many papers on principal
curve construction) ``softening'' starting from any linear
configuration of nodes does not lead to the expected solution. In
this case, adaptive strategies, like ``growing curve'' (an
analogue of what was used by K\'{e}gl in his polygonal algorithm
\cite{Kegl99} or ``growing surface'' can be used to obtain
suitable configuration of nodes.

We should mention the problem of good initial approximation for
$\phi$, used to initialize the algorithm. For principal manifolds,
one can start with the plane or the volume of several first
principal components calculated for $X$ (as in
\cite{4Bishop1998,Gorban01}), but other initializations could be
also developed.

Another important remark is the choice of elastic coefficients
$\mu$ and $\lambda$ such that they would give good approximation
to a hypothetical ``ideal'' principal manifold. The standard
strategies of using a training data subset or cross-validation
could be used. For example, we can remove a certain percentage of
points from the process of optimization and calculate
approximation error on this test set for every ``epoch" of
manifold softening, thus stopping with the best parameter values.
One should notice that despite the fact that we have two
parameters, the ``bending" elasticity parameter $\mu$ is more
important for optimizing this ``generalization" error, since it
determines the curvature of $\phi$. The stretching elasticity
parameter $\lambda$ is more important at the initial stage of net
optimization since it (together with $\mu$) forces regularity of
the node positions, making distances between neighboring nodes
almost uniform along the net. For the final stage of the
optimization process $\lambda$ can be even put to zero or some
small value.

\subsection{Missing Data Values}

Our geometrical approach to the construction of principal
manifolds allows us to deal very naturally with the \index{missing
data} missing values in the data. In bioinformatics this issue can
be very important: in one of the examples below, there are missing
values in almost any column and any row of the initial data table.
When filling gaps with mean values is not a good solution to the
problem, it is better to use principal objects for this purpose
(the mean value itself is the 0-dimensional principal object).

The general idea is to be able to project a data point $x_i$ with
 missing values in some of its coordinates onto the
principal object of the whole distribution $X$. In this case we
can not represent $x_i$ as a simple vector anymore. If $x_i$ has
$p$ missing values then it is natural to represent it as a linear
manifold parallel to the $p$ coordinate axes for which the data
values are missing: $x_i =
[x_i^0,...,t_1,...,x_i^k,...,t_p,..,x_i^m], -\infty \leq
t_1,t_2,...,t_p \geq \infty$. Then, for a principal manifold $y$,
we can calculate the closest distance (or the intersection point)
between two manifolds (the non-linear principal one and the linear
one representing $x_i$) and use the coordinate values of the
projected point. Theoretically, the reconstructed values are not
unique, when $x_i$ and $y$ intersect in many points or even along
lines, which gives several alternative solutions or even continuum
for the reconstructed missing values. With increasing data
dimension the probability of such intersections decreases.

\subsection{Adaptive Strategies}

The formulation of the method given above allows us to construct
different adaptive strategies by playing with a) individual
$\lambda _{i}$ and $\mu _{j}  $ weights; b) the grid connection
topology; c) the number of nodes. \index{adaptive strategies}

This is a way of extending the approach significantly making it
suitable for many practical applications.

First of all, let us define a basic operation on the grid, which
allows inserting new nodes. Let us denote by \textbf{N},
\textbf{S}, \textbf{R} the sets of all nodes, edges and ribs
respectively. Let us denote by \textbf{C}$(i)$ the set of all
nodes which are connected to the $i$th node by an edge. If one has
to insert a new node in the middle of an edge $I$, connecting two
nodes $k$ and $l$, then the following operations have to be
accomplished:

\vspace*{1mm}  \frame{
\begin{minipage}[l]{10.3cm}
\vspace{2mm}
\begin{enumerate}
\item{Delete from \textbf{R} those ribs which contain node $k$ or
node $l$;}
\item{Delete the edge $I  $ from \textbf{S};}
\item{Put a new node $m  $ in \textbf{N};}
\item{Put in \textbf{S} two new edges connecting $k$ and $m$, $m$ and
$l$;}
\item{Put in \textbf{R} new ribs, connecting $m$, $k$ and all
$i \in $\textbf{C}($k)$, and $m$, $l$ and all $i \in
$\textbf{C}($l)$.}
\end{enumerate}
\vspace{0.01mm}
\end{minipage}
}\vspace*{1mm}\\At steps 4, 5 one has to assign new weights to the
edges and ribs. This choice depends on the task to be solved. If
one constructs a ``growing'' grid, then these weights must be
chosen to be the same as they were for the deleted ones. If one
constructs a refinement of an already constructed grid, one must
choose these weights to be twice bigger than they were for the
deleted ones.

The \textit{grow-type strategy} is applicable mainly to grids with
planar topology (linear, rectangular, cubic grids). It consists of
an iterative determination of those grid parts, which have the
largest ``load'' and doubling the number of nodes in this part of
the grid. The load can be defined in different ways. One natural
way is to calculate the number of points that are projected onto
the nodes. For linear grids the grow-type strategy consists of the
following steps:

\vspace*{1mm}
 \frame{
\begin{minipage}[l]{10.3cm}
\vspace{2mm}

\begin{enumerate}
\item{Initializing the grid; it must contain at least two nodes and
one edge;}
\item{Determining the edge which has the largest load, by
summing the number of data points (or the sum of their weights)
projected to both ends of every edge;}
\item{Inserting a new node in
the middle of the edge, following the operations described above;}
\item{Optimizing the positions of the nodes.}
\end{enumerate} \vspace{0.01mm}
\end{minipage}
}\vspace*{1mm} \\ One stops this process usually when a certain
number of nodes in the grid is reached (see, for example,
\cite{Kegl00}). This number is connected with the total amount of
points. In the {\textit{elmap}} package \cite{Elmap} this is an
explicit parameter of the method, allowing a user to implement his
own stopping criterion. Because of this stopping condition the
computational complexity is not proportional to the number of data
points and, for example, grows like $n^{{\rm 5}{\rm /} {\rm 3}}  $
in the case of the Polygonal Line algorithm. Another form of the
stopping condition is when the mean-square error (MSE) does not
change more than a small number $\varepsilon $ after several
insertion/optimization operations.

We should mention here also \index{growing lump} {\it growing
lump} and {\it growing flag} \index{growing flag} strategies used
in physical and chemical applications \cite{Grids,CMIM}. In the
growing lump strategy we add new nodes uniformly at the boundary
of the grid using a linear extrapolation of the grid embedding.
Then the optimization step follows, and, after that, again the
step of growing could be done.

For the principal flag \index{principal flag} one uses
sufficiently regular grids, in which many points are situated on
the coordinate lines, planes, etc. First, we build a
one-dimensional grid (as a one-dimensional growing lump, for
example). Then we add a new coordinate and start growing in new
direction by adding lines. After that, we can add the third
coordinate, and so on.

The \textit{break}-type adaptive strategy \index{break-type
adaptive strategy} changes individual rib weights in order to
adapt the grid to those regions of data space where the
``curvature'' of data distribution has a break or is very
different from the average value. It is particularly useful in
applications of principal curves for contour extraction (see
\cite{GorbanZinovyev05Computing}). For this purpose the following
steps are performed:

\vspace*{1mm} \frame{
\begin{minipage}[l]{10.3cm}
\vspace{2mm}

\begin{enumerate}
\item{Collect statistics for the distances from every node $i$ to
the mean point of the datapoints that are projected into this node:
$$r_{j} = {\left\| {\phi(y_{j}) - \left(\sum\limits_{x^{(i)} \in K^{y_j}}
{w_{i}}\right)^{-1} {\sum\limits_{x^{(i)} \in K^{y_j}} ^{} {w_{i}
x^{(i)}}}} \right\|}.$$}
\item{Calculate the mean value and the standard deviation for some power of
{\textit{r}} : $m = \overline {r^{\alpha}}  $, $s = \sigma
_{r^{\alpha}}  $; where $\alpha>1$ is a parameter which in our
experiments is chosen to be 4.}
\item{Determine those nodes for
which $r_{j} > m+\beta s$, where $\beta > 0$ is another parameter,
equal 2 in our experiments.}
\item{For every node $k$ determined
at the previous step one finds those ribs that have $k$ as their
central point and change their weight for $\mu _{j}^{(new)} = \mu
_{j}^{(old)} \cdot {\frac{{m}}{{r_{j}^{\alpha}} } }$.}
\item{Optimize the node positions.}
\item{Repeat this process a certain number of times.}
\end{enumerate}
\vspace{0.01mm}
\end{minipage}
}\vspace*{1mm} \\ The \textit{principal graph} strategy,
implemented in the {\textit{elmap}} package \cite{Elmap} allows us
to perform clustering of curvilinear data features along principal
curves. Two example applications of this approach are satellite
image analysis \cite{Banfield92} or hand-written symbol
skeletonization \cite{Kegl02,GorbanZinovyev05Computing}. First,
notice that the grid we constructed does not have to be a
connected graph. The system matrix (6) is not singular if for
every connected component of the graph there are data points that
are projected onto one of its nodes. This allows using the same
algorithmic kernel to optimize node positions of the unconnected
graph. Notice also that if the sets of edges and ribs are empty,
then this algorithm acts exactly like the standard $K$-means
clustering algorithm.

To construct a ``skeleton'' for a two-dimensional point
distribution, we apply a variant of local linear principal
\index{skeleton} component analysis first, then connect local
components into several connected parts and optimize the node
positions afterwards. This procedure is robust and efficient in
applications to clustering along curvilinear features and it was
implemented as a part of the {\textit{elmap}} package
\cite{Elmap}. The following steps are performed:

\vspace*{1mm} \frame{
\begin{minipage}[l]{10.3cm}
\vspace{2mm}
\begin{enumerate}
\item{Make a ``grid'' from a number of unconnected nodes (sets of edges
and ribs are empty at this stage). Optimize the node positions
(i.e., do $K$-means clustering for large $k$). The number of nodes
$k$ is chosen to be a certain proportion of the number of data
points. In our experiments we used $k=5\%$ of the total number of
data points.}
\item{For
every node of the grid in position $y^{i}$, the local first
principal direction is calculated. By local we mean that the
principal direction is calculated inside the cluster of datapoints
corresponding to the node $i$. Then this node is substituted by two
new nodes in positions $y^{(new1)}= y^{i}+\alpha
s\textbf{\textit{n}}, y^{ (new2)}= y^{i} - \alpha s
\textbf{\textit{n}}$, where \textbf{\textit{n}} is the unit vector
in the principal direction, $s$ is the standard deviation of data
points belonging to the node $i$, $\alpha $ is a parameter, which
can be taken to be 1.}
\item{A collection of edges and ribs is generated,
following this simple rule: every node is connected to the node
which is closest to this node but not already connected at step 2,
and every such connection generates two ribs, consisting of a new
edge and one of the edges made at step 2.}
\item{Weights
of the ribs are calculated. A rib is assigned a weight equal to
$\vert $cos($\alpha )\vert $, where $\alpha $ is an intersection
angle of two edges contained in this rib, if $\alpha   \ge
\frac{\pi }{2} $. Otherwise it is zero (or, equally, the rib is
eliminated).}
\item{The node positions are optimized.}
\end{enumerate}
\vspace{0.01mm}
\end{minipage}
}\vspace*{1mm} \\ One possible way to improve the resulting graph
further is to apply graph simplification rules, analogously to the
method in \cite{Kegl02}. The idea of this algorithm is close to
the $k$-segments algorithm of Verbeek \cite{Verbeek00} and,
indeed, one possible option is to use $k$-segment clustering
instead of $K$-means clustering on the first step of the
algorithm.

The adaptive strategies: ``grow'', ``break'' and ``principal''
graphs can be combined and applied one after another. For example,
the principal graph strategy can be followed by break-type weight
adaptation or by grow-type grid adaptation.

\section{Elastic Maps}

\subsection{Piecewise Linear Manifolds and Data Projectors}\label{projectionsection}

\index{elastic map}In the process of net optimization we use
projection of data into the closest node. This allows us to
improve the speed of the data projection step without loosing too
much detail when the grid resolution is good enough. The effect of
an estimation bias, connected with this type of projection, was
observed in \cite{KeglThesis99}. In our approach the bias is
indirectly reduced by utilizing the $U_E^\phi$ term that makes the
grid almost isometric (having the same form, the grid will have
lesser energy with equal edge lengths). For presentation of data
points or for data compression, other projectors can be applied.

In applications we utilize a piece-wise linear projection onto the
manifold. A natural way to do this is to introduce a set of
simplexes on the grid (line segments for one-dimensional grids,
triangles for two-dimensional grids, and tetrahedra for the 3D
grids). Then one performs orthogonal projection onto this set. In
order to avoid calculation of all distances to all simplexes, one
can apply a simplified version of the projector: find the closest
node of the grid and then consider only those simplexes that
contain this node. This type of projection is used in the examples
(Fig.~\ref{DatasetI}-\ref{DatasetIII}).

Since the elastic net has a \index{penalty} penalty on its length
(and, for higher dimensions, indirectly, area, volume), the result
of the optimization procedure is a bounded manifold, embedded in
the cloud of data points. Because of this, if the penalty
coefficient is big, many points can have projection on the
boundary of the manifold. This can be undesirable, for example, in
data visualization applications. To avoid this effect, a linear
extrapolation of the bounded rectangular manifold (extending it by
continuity in all directions) is used as a post-processing step.

Using piece-wise linear interpolation between nodes and linear
extrapolation of the manifold is the simplest (hence
computationally effective) possibility. It is equivalent to
construction of piece-wise linear manifold. Other interpolation
and extrapolation methods are also applicable, for example the use
of Carleman's formula
\cite{Aizenberg93,Grids,CMIM,Gorban02,GapsGRW,GapsDGRKM} or using
Lagrange polynomials \cite{Ritter93}, although they are more
computationally expensive.

\subsection{Iterative Data Approximation}

An important limitation of many approaches that use grid
approximations is that in practice it is feasible to construct only
low-dimensional grids (no more than three-dimensional).

This limitation can be overcome in the additive iterative manner.
After low-dimensional manifold $\phi$ is constructed, for every
point $x$, its projection $P^\phi(x)$ onto the manifold is
subtracted and new low-dimensional manifold is computed for the
set of residues $\{x_i-P^\phi(x_i)\}$. Thus a dataset can be
approximated by a number of low-dimensional objects, each of them
approximating the residues obtained from the approximation by the
previous one. The norm of the residues at each iteration becomes
smaller but is not guaranteed to become zero after a finite number
of iterations, as in the case of linear manifolds.

\section{Principal Manifold as Elastic Membrane}

Let us discuss in more detail the central idea of our approach:
using the metaphor of elastic membranes in the principal manifold
construction algorithm. The system represented in Fig.~\ref{4Fig3}
can be modeled as an elastic membrane with external forces applied
to the nodes. In this section we consider the question of
equivalence between our spring network system and realistic physical
systems (evidently, we make comparison in 3D).

Spring meshes are widely used to create physical models of elastic
media (for example, \cite{Born54}). The advantages in comparison
with the continuous approaches like the Finite Elements Method
(FEM), are evident: computational speed, flexibility, the
possibility to solve the inverse elasticity problem easily
\cite{VanGelder97}.

Modeling elastic media by spring networks has a number of
applications in computer graphics, where, for example, there is a
need to create realistic models of soft tissues (human skin, as an
example). In \cite{VanGelder97} it was shown that it is not
generally possible to model elastic behavior of a membrane using
spring meshes with simple scalar springs. In \cite{Xie02} the
authors introduced complex system of penalizing terms to take into
account angles between scalar springs as well as shear elasticity
terms. This improved the results of modeling and has found
applications in subdivision surface design.

In a recent paper \cite{Gusev04} a simple but important fact was
noticed: every system of elastic finite elements could be
represented by a system of springs, if we allow some springs to
have negative elasticity coefficients. The energy of a $k$-star
$S_k$ in $R^m$ with $y_0$ in the center and $k$ endpoints
$y_{1,...,k}$ is $u_{S_k} = \mu_{S_k}(\sum_{i=1}^ky_i-k y_0)^2$,
or, in the spring representation, $u_{S_k} =
k\mu_{S_k}\sum_{i=1}^k(y_i-y_0)^2-\mu_{S_k}\sum_{i>j}(y_i-y_j)^2$.
Here we have $k$ positive springs with coefficients $k\mu_{S_k}$
and $k(k-1)/2$ negative springs with coefficients $-\mu_{S_k}$.

Let us slightly reformulate our problem to make it closer to the
standard notations in the elasticity theory. We introduce the $m
\times p$-dimensional vector of displacements, stacking all
coordinates for every node:

\[
u = \{u_{1}^{(1)} ;u_{2}^{(1)} ;..;u_{m}^{(1)} ;...;u_{1}^{(p)}
;u_{2}^{(p)} ;..;u_{m}^{(p)} \}^{T}\; ,\]

\noindent where $m$ is dimension, $p$ is the number of nodes,
$u_{i}^{(k)}  $ is the $i$th component of the $k$th node
displacement. The absolute positions of nodes are $y_{}^{(k)} =
\tilde {y}_{}^{(k)} + u_{}^{(k)} $, where $\tilde {y}_{}^{(k)} $
are equilibrium (relaxed) positions. Then our minimization problem
can be stated in the following generalized form:

\begin{equation}
\label{4eq6} \quad u^{T}Eu + D(u;x) \to \min \; ,
\end{equation}

\noindent where $E  $ is a symmetric $(m \times p) \times (m \times
p)$ element \index{stiffness matrix} stiffness matrix. This matrix
reflects the elastic properties of the spring network and has the
following properties: 1) it is sparse; 2) it is invariant with
respect to translations of the whole system (as a result, for any
band of $m$ consecutive rows corresponding to a given node $k$, the
sum of the m$m \times m$ off-diagonal blocks should always be equal
to the corresponding diagonal block taken with the opposite sign).
The $D(u;x)$ term describes how well the set of data $x  $ is
approximated by the spring network with the node displacement vector
$u$. It can be interpreted as the energy of external forces applied
to the nodes of the system. To minimize (\ref{4eq6}) we solve the
problem of finding equilibrium between the elastic internal forces
of the system (defined by $E$) and the external forces:

\begin{equation}
\label{4eq7} Eu = f,\;\;\;f = -
{\frac{{1}}{{2}}}{\frac{{\partial}} {{\partial u}}}D(u;x)\; .
\end{equation}

The matrix $E$ is assembled in a similar manner to that described in
Subsec.~\ref{subsec:basOptAlg}. There is one important point: the
springs have zero rest lengths, it means that equilibrium node
positions are all in zero: $\tilde {y}_{}^{(k)} = 0,\;k = 1..p$. The
system behavior then can be described as ``super-elastic''. From the
point of view of data analysis it means that we do not impose any
pre-defined shape on the data cloud structure.

For $D(u$;$x)$ we use the usual mean square distance measure, see
(\ref{approxerror}): $D(u;x) = U^\phi_A$. The force applied to the
$j$th node equals

\begin{equation}
\label{4eq8} f_{j} = {\frac{{n^{(j)}}}{{N}}}\left( {\tilde
{x}^{(j)} - u^{(j)}} \right)\;,
\end{equation}

\noindent where

\[
\tilde {x}^{(j)} = {\frac{{{\sum\limits_{x^{(i)} \in K^{y_j}}^{}
{w_{i} x^{(i)}}}} }{{n^{(j)}}}}\; , \;\;n^{(j)} =
{\sum\limits_{x^{(i)} \in K^{y_j}} {w_{i}}} \;  ,\;\;\;\;N =
{\sum\limits_{x^{(i)}} {w_{i}}}\; .
\]

It is proportional to the vector connecting the $j$th node and the
weighted average $\tilde {x}^{(j)} $ of the data points in $K^{{\rm
(}j{\rm )}}_{ } $ (i.e., the average of the points that surround the
$j$th node: see (\ref{taxondef}) for definition of $K^{y_j})$. The
proportionality factor is simply the relative size of $K^{y_j}$. The
linear structure of (\ref{4eq8}) allows us to move $u$ in the left
part of the equation (\ref{4eq8}). Thus the problem is linear.

Now let us show how we can benefit from the definition (\ref{4eq7})
of the problem. First, we can introduce a pre-defined equilibrium
shape of the manifold: this initial shape will be elastically
deformed to fit the data. This approach corresponds to introducing a
model into the data. Secondly, one can try to change the form
(\ref{4eq8}) of the external forces applied to the system. In this
way one can utilize other, more sophisticated approximation
measures: for example, taking outliers into account.

Third, in three-dimensional applications one can benefit from
existing solvers for finding equilibrium forms of elastic
membranes. For multidimensional data point distributions one has
to adapt them, but this adaptation is mostly formal.

Finally, there is a possibility of a hybrid approach: first, we
utilize a ``super-elastic'' energy functional to find the initial
approximation. Then we ``fix'' the result and define it as the
equilibrium. Finally, we utilize a physical elastic functional to
find elastic deformation of the equilibrium form to fit the data.

\section{Method Implementation}

For different applications we made several implementations of the
{\it elmap} algorithm in the C++, Java and Delphi programming
languages. All of them are available on-line
\cite{Elmap,VidaExpert,VIMIDA} or from the authors by request.

Two graphical user interfaces for constructing elastic maps have
also been implemented. The first one is the VidaExpert stand-alone
application \cite{VidaExpert} which provides the possibility not
only to construct elastic maps of various topologies but also to
utilize linear and non-linear principal manifolds to visualize the
results of other classical methods for data analysis (clustering,
classification, etc.) VidaExpert has integrated 3D-viewer which
allows data manipulations in interactive manner.

A Java implementation was also created \cite{VIMIDA}, it was used
to create a Java-applet {\it ViMiDa} (VIsualization of MIcroarray
DAta). In the Bioinformatics Laboratory of Institut Curie (Paris)
it is integrated with other high-throughput data analysis
pipelines and serves as a data visualization user front-end
(``one-click" solution). ViMiDa allows remote datasets to be
downloaded and their visualization using linear or non-linear
(elastic maps) principal manifolds.

\section{Examples}

\subsection{Test Examples}

\begin{figure}[t]
\centering{ \textbf{a}) spiral \includegraphics[width=43mm,
height=43mm]{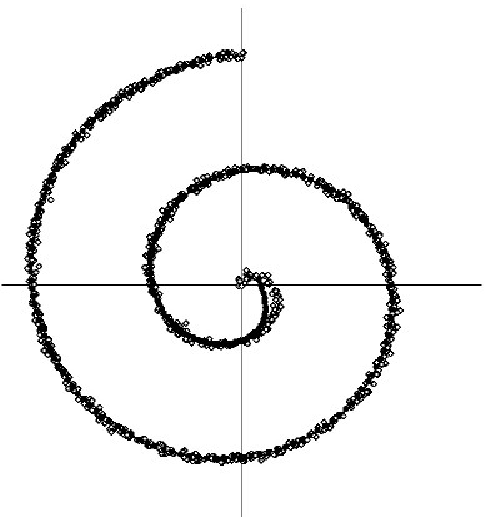} \textbf{b}) large
\includegraphics[width=43mm, height=43mm]{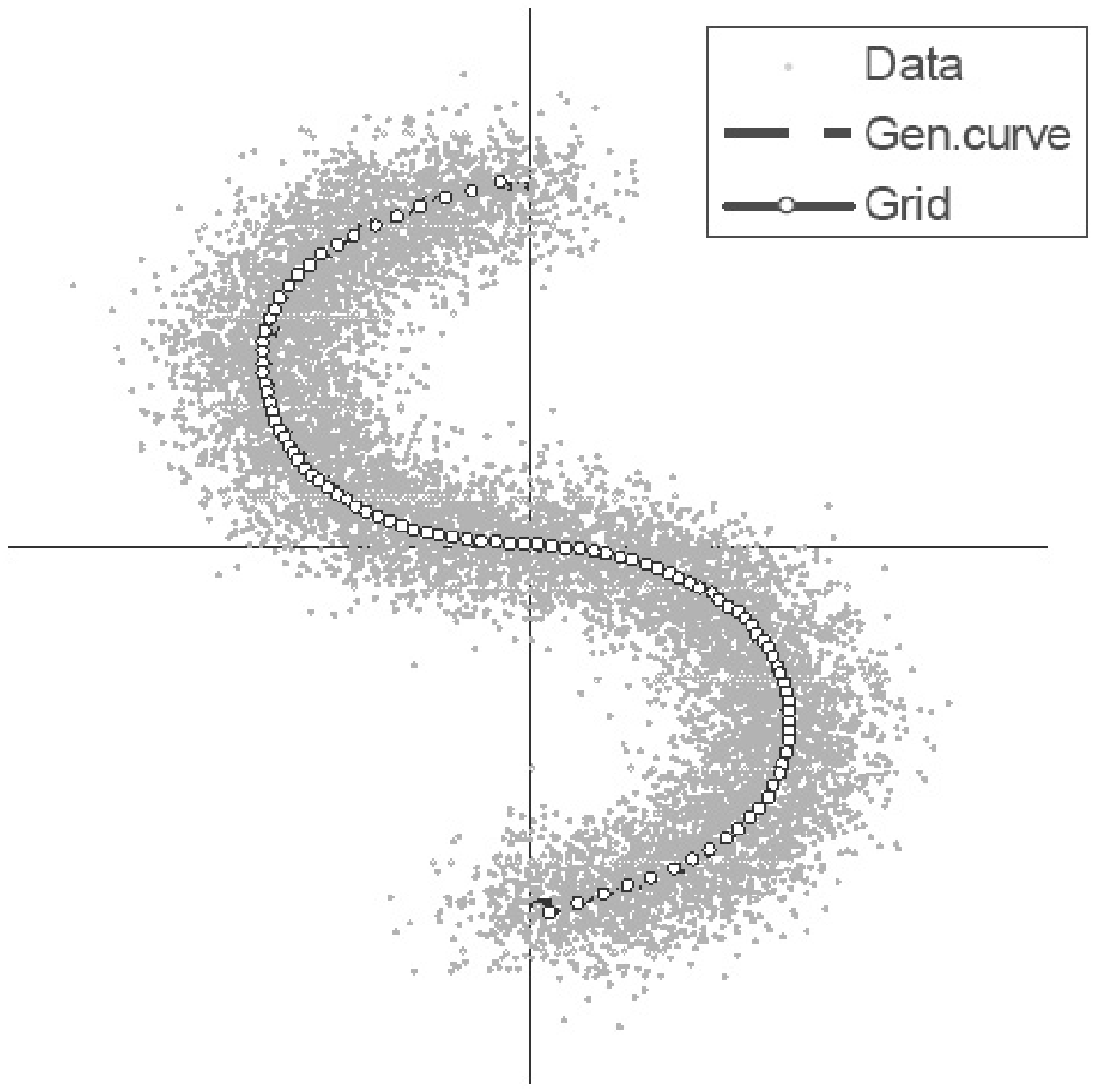} }
\caption{Two-dimensional examples of principal curves construction
\label{4Fig6}}
\end{figure}

In Fig.~\ref{4Fig6} we present two examples of 2D-datasets provided
by K\'{e}gl\footnote{http://www.iro.umontreal.ca/ $ \sim
 $ Kegl/research/pcurves/implementations\\ /Samples/}.

The first dataset called \textit{spiral} is one of the standard
ways in the principal curve literature to show that one's approach
has better performance than the initial algorithm provided by
Hastie and Stuetzle. As we have already mentioned, this is a bad
case for optimization strategies, which start from a linear
distribution of nodes and try to optimize all the nodes together
in one loop. But the adaptive ``growing curve'' strategy, though
being by order of magnitude slower than ``softening'', finds the
solution quite stably, with the exception of the region in the
neighborhood of zero, where the spiral has very different
(compared to the average) curvature.

The second dataset, called \textit{large} presents a simple
situation, despite the fact that it has comparatively large sample
size (10000 points). The nature of this simplicity lies in the fact
that the initial first principal component based approximation is
already effective; the distribution is in fact \textit{quasilinear},
since the principal curve can be unambiguously orthogonally
projected onto a line. In Fig.~\ref{4Fig6}b it is shown that the
generating curve, which was used to produce this dataset, has been
discovered almost perfectly and in a very short time.

Some other examples, in particular of application of the {\it
elastic graph} \hspace{5mm} strategy are provided in
\cite{GorbanZinovyev05Computing}.

\subsection{Modeling Molecular Surfaces}

A molecular surface defines an effective region of space which is
occupied by a molecule. For example, the Van-der-Waals molecular
surface \index{molecular surface} is formed by surrounding every
atom in the molecule by a sphere of radius equal to the
characteristic radius of the Van-der-Waals force. After all the
interior points are eliminated, this forms a complicated
non-smooth surface in 3D. In practice, this surface is sampled by
a finite number of points.

Using principal manifolds methodology, we constructed a smooth
approximation of such molecular surface for a small piece of a DNA
molecule (several nucleotides long). For this we used slightly
modified Rasmol source code \cite{Sayle92} (available from the
authors by request).

\begin{figure}[t]
\centering{
\includegraphics[width=50mm,
height=50mm]{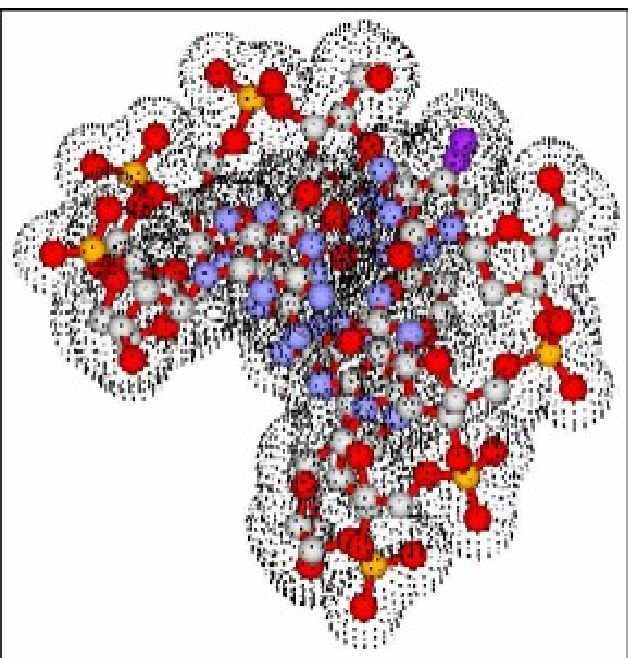}
\includegraphics[width=50mm,
height=50mm]{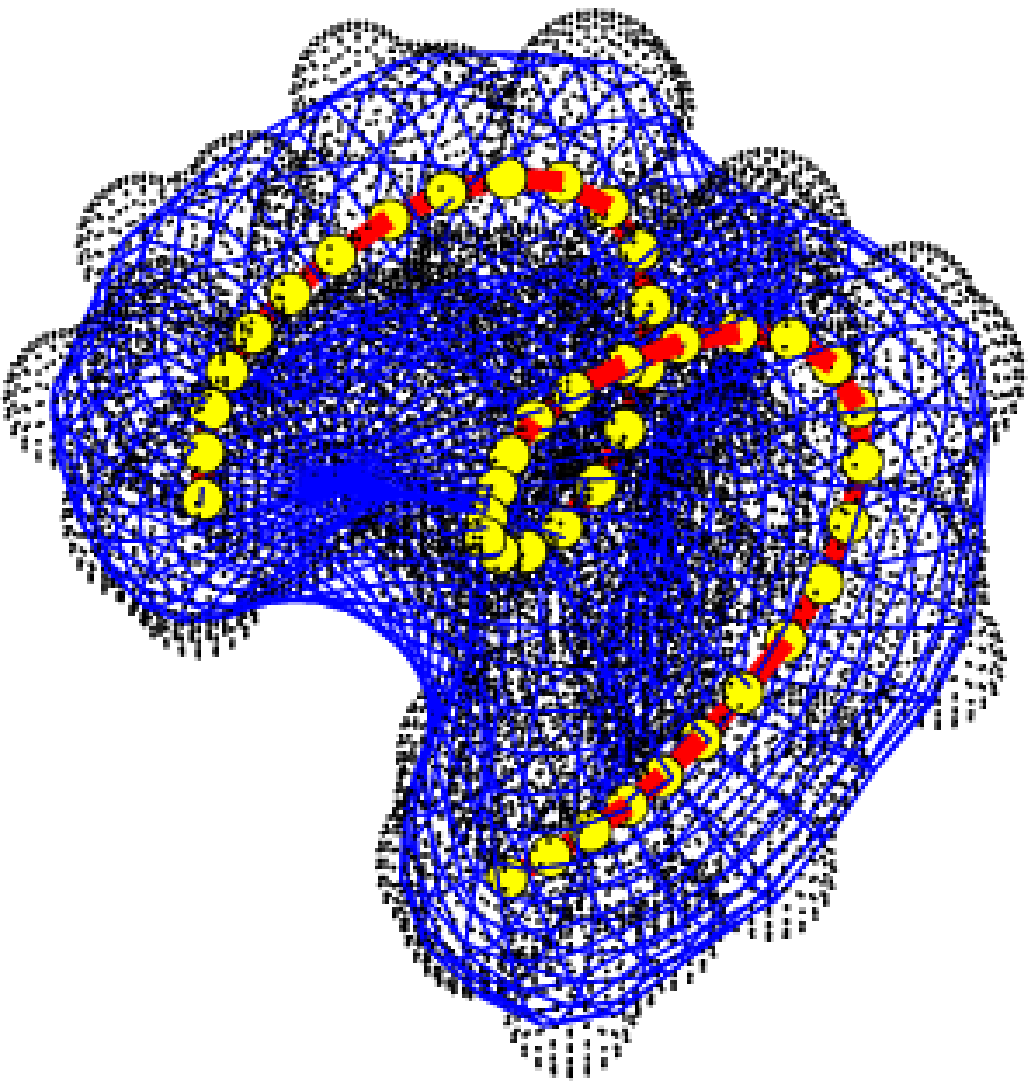}\caption{Approximating molecular surface of
a simple fragment of DNA by 1D- and 2D- principal manifolds. On
the left the molecule with molecular surface is shown, on the
right the result of the approximation is shown. The small points
are the points on the molecular surface, the big spiral is the 1D
principal curve and the mesh is the 2D principal spherical
manifold} \label{Fig7}}
\end{figure}

First, we have made an approximation of this dataset by a 1D
principal curve. Interestingly, this curve followed the backbone
of the molecule, forming a spiral (see Fig.~\ref{Fig7}).

Second, we approximated the molecular surface by a 2D manifold.
The topology of the surface is expected to be spherical, so we
applied spherical topology of the elastic net for fitting. We
should notice that since it is impossible to make the lengths of
all edges equal for the sphere-like grid, corrections were
performed for edge and rib weights during the grid initialization
(shorter edges are given with larger weights proportionally and
the same for the ribs).

As a result one gets a smooth principal manifold with spherical
topology, approximating a rather complicated set of points. The
result of applying principal manifold construction using the {\it
elmap} \cite{Elmap} package is shown in Fig.~\ref{Fig7}. The
manifold allows us to introduce a global spherical coordinate
system on the molecular surface. The advantage of this method is
its ability to deal not only with star-like shapes as the
spherical harmonic functions approach does (see, for example,
\cite{Cai02}) but also to model complex forms with cavities as
well as non-spherical forms.

\subsection{Visualization of Microarray Data}

DNA microarray data is a rich source of information for molecular
biology (for a recent overview, see \cite{Leung2003}). This
technology found numerous applications in understanding various
biological processes including cancer. It allows to screen
simultaneously the expression of all genes in a cell exposed to
some specific conditions (for example, stress, cancer, normal
conditions). Obtaining a sufficient number of observations
(chips), one can construct a table of ``samples vs genes",
containing logarithms of the expression levels of, typically
several thousands ($n$) of genes, in typically several tens ($m$)
of samples.

The table can be represented as a set of points in two different
spaces. Let us call them the ``gene space" and the ``sample
space". In the gene space every point (vector of dimension $m$) is
a gene, characterized by its expression in $m$ samples. In the
sample space every point is a sample (vector of dimension $n$),
characterized by the expression profile of $n$ genes.

One of the basic features of \index{microarray}microarray datasets
is the fact that $n\gg~m$. This creates certain problems when one
constructs a classifier in the ``sample space": for example, for
standard Linear Discriminant Analysis, too many equivalent
solutions for the problem are possible. Thus, some regularization
of the classification problem is required (see, for example,
\cite{Hastie05}) and PCA-like techniques could be also used for
such a purpose.

Here we provide results on the visualization of three microarray
datasets in the ``sample space" using elastic maps. We demonstrate
that the two-dimensional principal manifolds outperform the
corresponding two-dimen\-sio\-nal principal planes, in terms of
representation of big and small distances in the initial
multidimensional space and visualizing class information.

\subsubsection{Datasets used}

For the participants of the international workshop ``Principal
manifolds-2006" which took place in Leicester, UK, in August of
2006, it was proposed to analyze three datasets containing
published results of microarray technology application. These
datasets can be downloaded from the workshop web-page
\cite{PM2006Webpage}.

\noindent {\it Dataset I. Breast cancer microarrays.}
\index{cancer}This dataset was published in \cite{Wang2005} and it
contains 286 samples and expression values of 17816 genes. We
processed the clinical information provided with this dataset and
constructed three {\it ab initio} sample classifications:

GROUP: non-agressive (A) vs agressive (B) cancer;

ER: estrogen-receptor positive (ER+) vs estrogen-receptor negative
(ER-) tumors;

TYPE: sub-typing of breast cancer (lumA, lumB, normal, errb2, basal
and unclassified), based on the so-called Sorlie breast cancer
molecular portraits (see \cite{Sorlie2000}).

\noindent {\it Dataset II. Bladder cancer microarrays.} This dataset
published in \cite{Dyrskjot03} contains 40 samples of bladder
tumors. There are two {\it ab initio} sample classficitions:

TYPE: clinical classification of tumors (T1, T2+ anf Ta);

GRADE: tumor grade (2,3,4).

\noindent {\it Dataset III. Collection of microarrays for normal
human tissues.} \index{tissue} This dataset published in
\cite{Shyamsundar2005} contains gene expression values in 103
samples of normal human tissues. The only {\it ab initio} sample
classification corresponds to the tissue type from which the sample
was taken.

Microarray datasets are inevitably incomplete: for example, there
is a number of gaps (not reliably measured values) in Dataset III.
Datasets I and II do not contain missing values (in Dataset I the
missing values are recovered with use of some standard
normalization method and in the Dataset II all gaps are filtered
out). Dataset III is filtered in such a way that any row (gene)
has at least 75\% complete values of expression and every column
has at least 80\% of complete values. It is interesting to mention
that missing values in Dataset III are distributed in such a way
that almost every column and row contains a missing value. This
means that missing value recovery is absolutely necessary for its
analysis.

The original expression tables were processed to select 3000 genes
with the most variable expression between samples. Thus the sample
space has approximately equal dimensions (3000) for all three
datasets. For every gene, the expression values were converted to
$z$-values, i.e. every value was centered and divided by its
standard deviation.

\begin{figure}
\centering{ \includegraphics[width=110mm,
height=75mm]{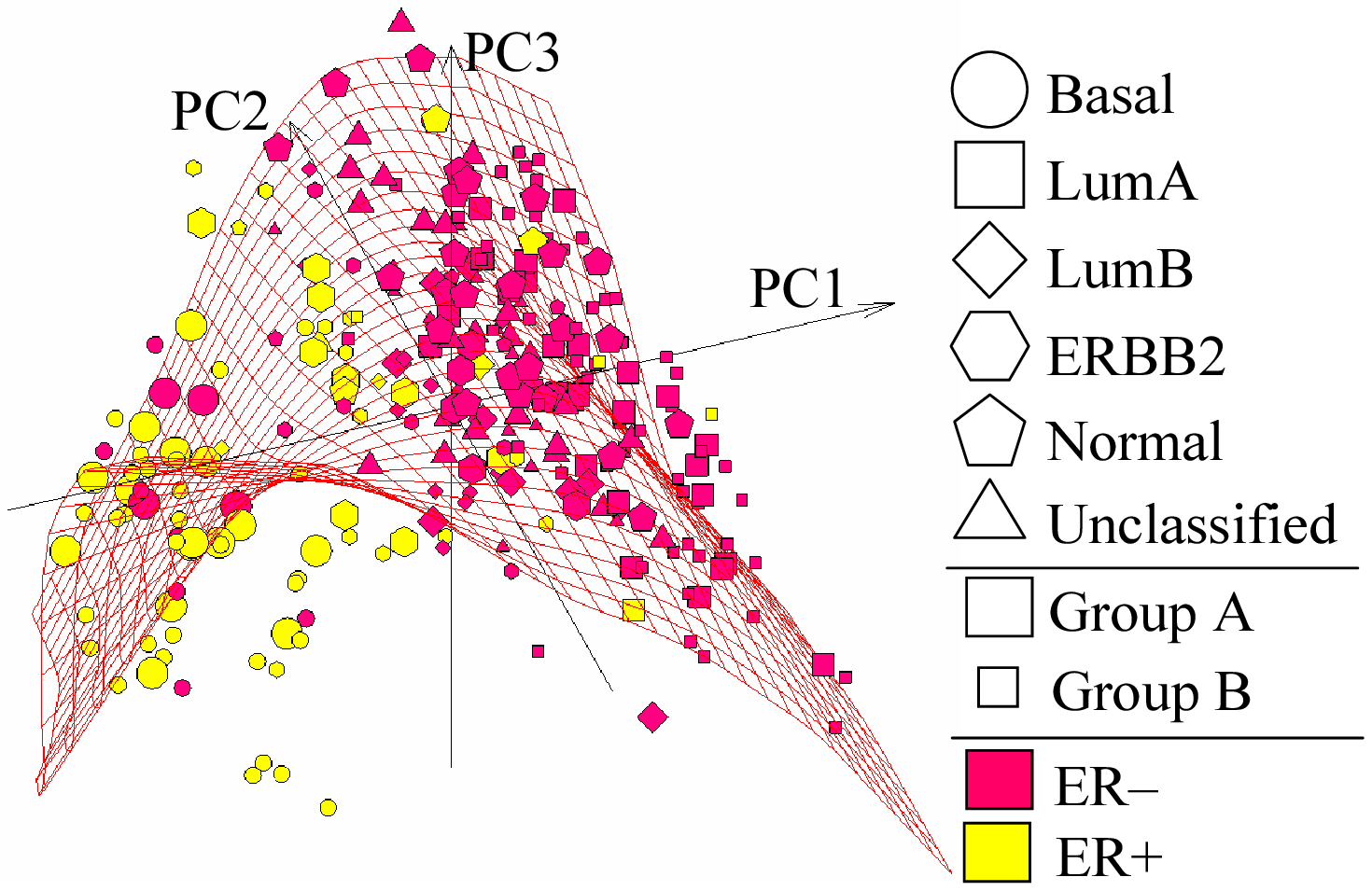}\\ \hspace{0mm} \textbf{a}) \hspace{60mm}
\\
\includegraphics[width=55mm, height=55mm]{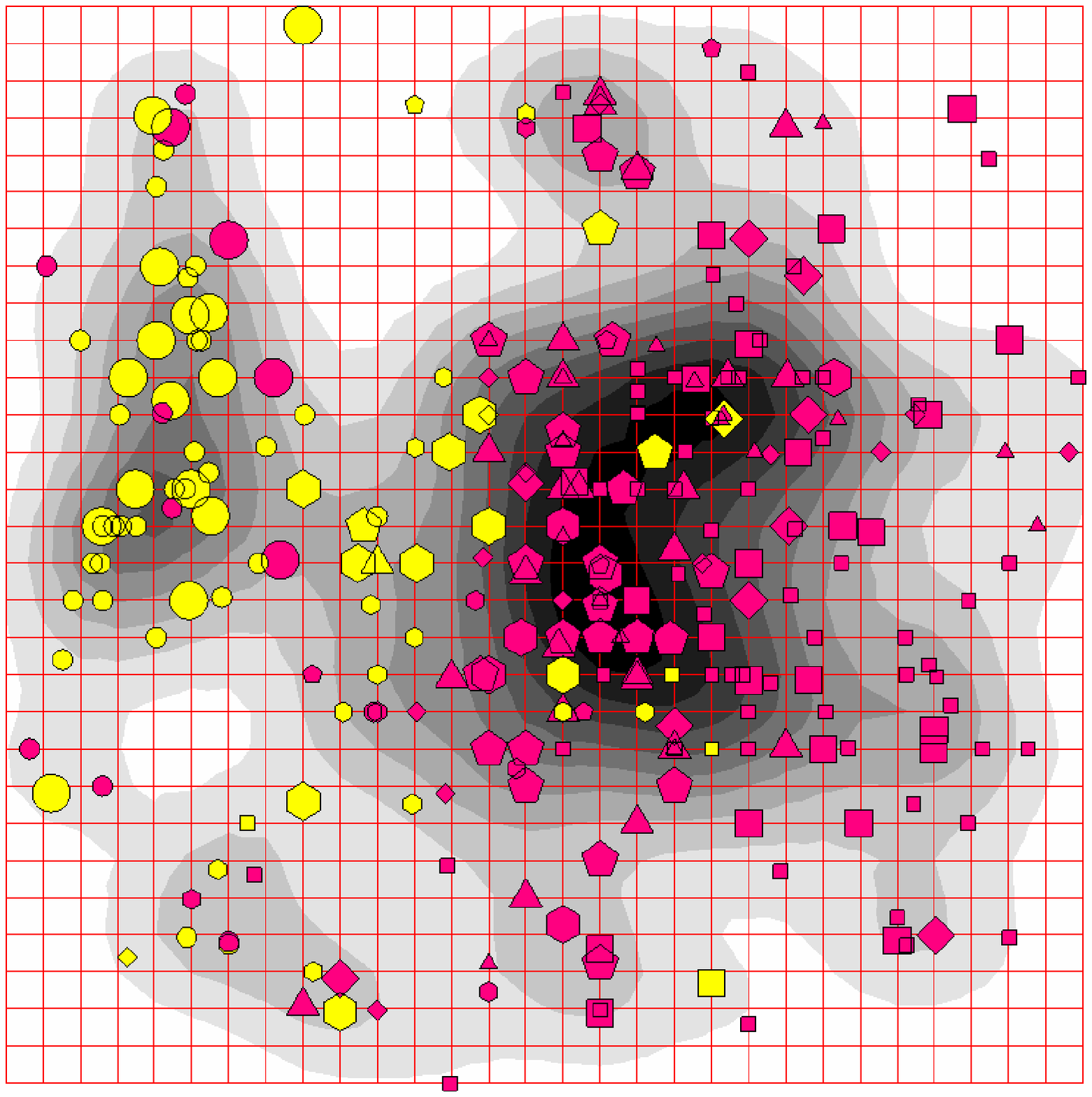}
\includegraphics[width=55mm, height=55mm]{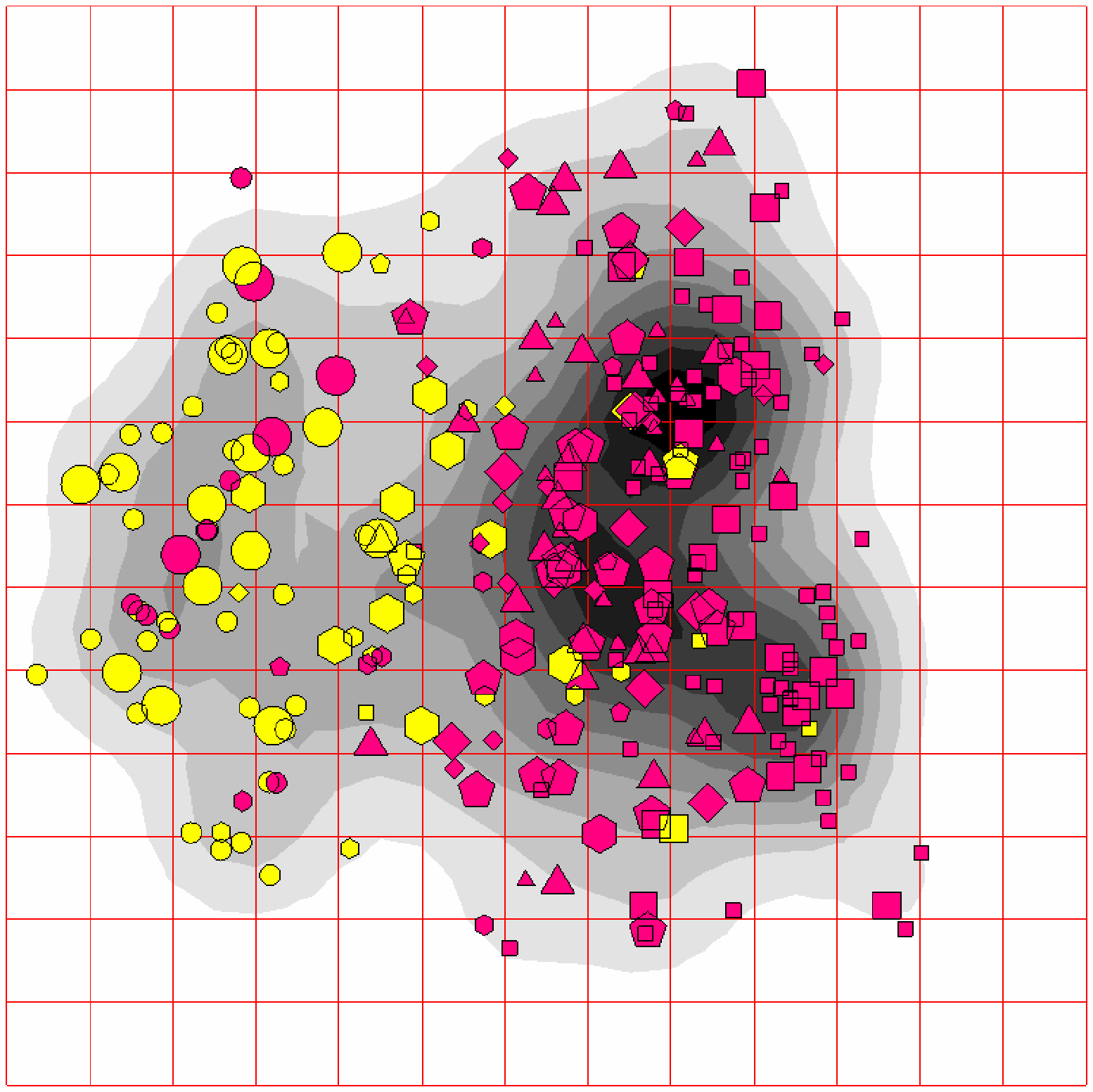}\\
\textbf{b}) ELMap2D \hspace{35mm} \textbf{c}) PCA2D
 } \caption{Visualization of Dataset I (breast cancer dataset)
using elastic maps. {\it Ab initio} classifications are shown
using points size (ER), shape (GROUP) and color (TYPE):
\textbf{a}) configuration of nodes in the three-dimensional
principal linear manifold. One clear feature is that the dataset
is curved such that it can not be mapped adequately on a
two-dimensional principal plane; \textbf{b}) the distribution of
points in the internal non-linear manifold coordinates (ELMap2D)
is shown together with an estimation of the two-dimensional
density of points; \textbf{c}) the same as b), but for the linear
two-dimensional manifold (PCA2D). One can notice that the
``basal'' breast cancer subtype is visualized more adequately with
ELMap2D and some features of the distribution become better
resolved in comparison to PCA2D \label{DatasetI}}
\end{figure}

\begin{figure}
\centering{ \includegraphics[width=110mm, height=75mm]{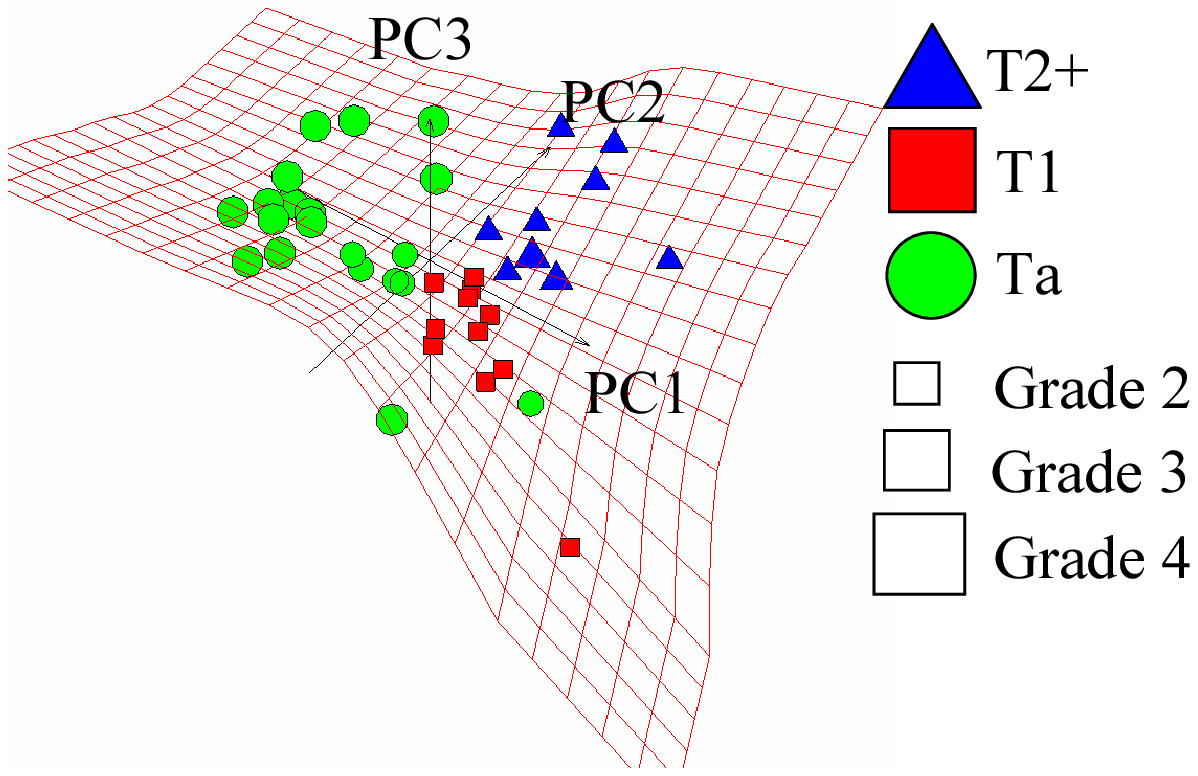}\\
\hspace{0mm} \textbf{a}) \hspace{60mm} \\
\includegraphics[width=55mm, height=55mm]{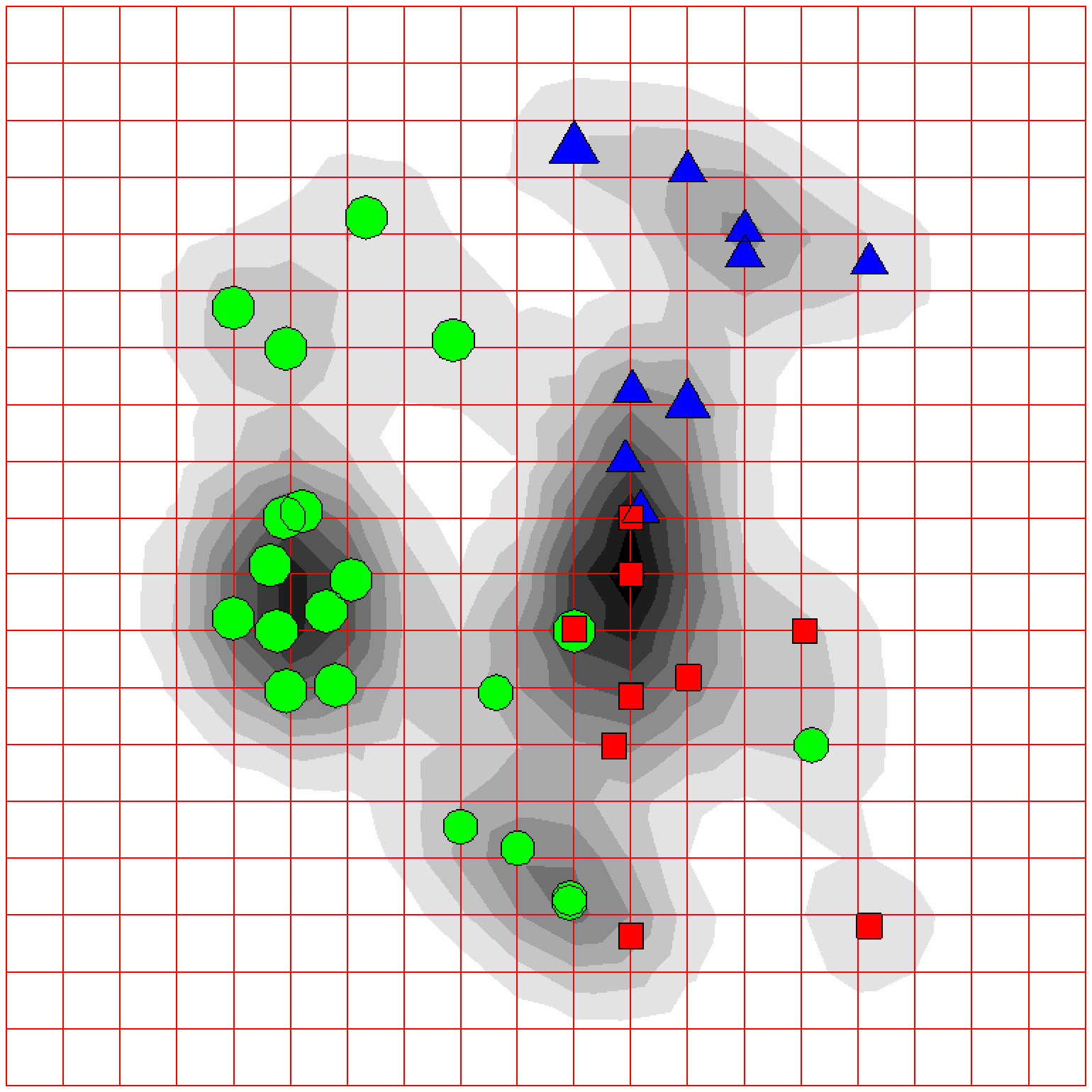}
\includegraphics[width=55mm, height=55mm]{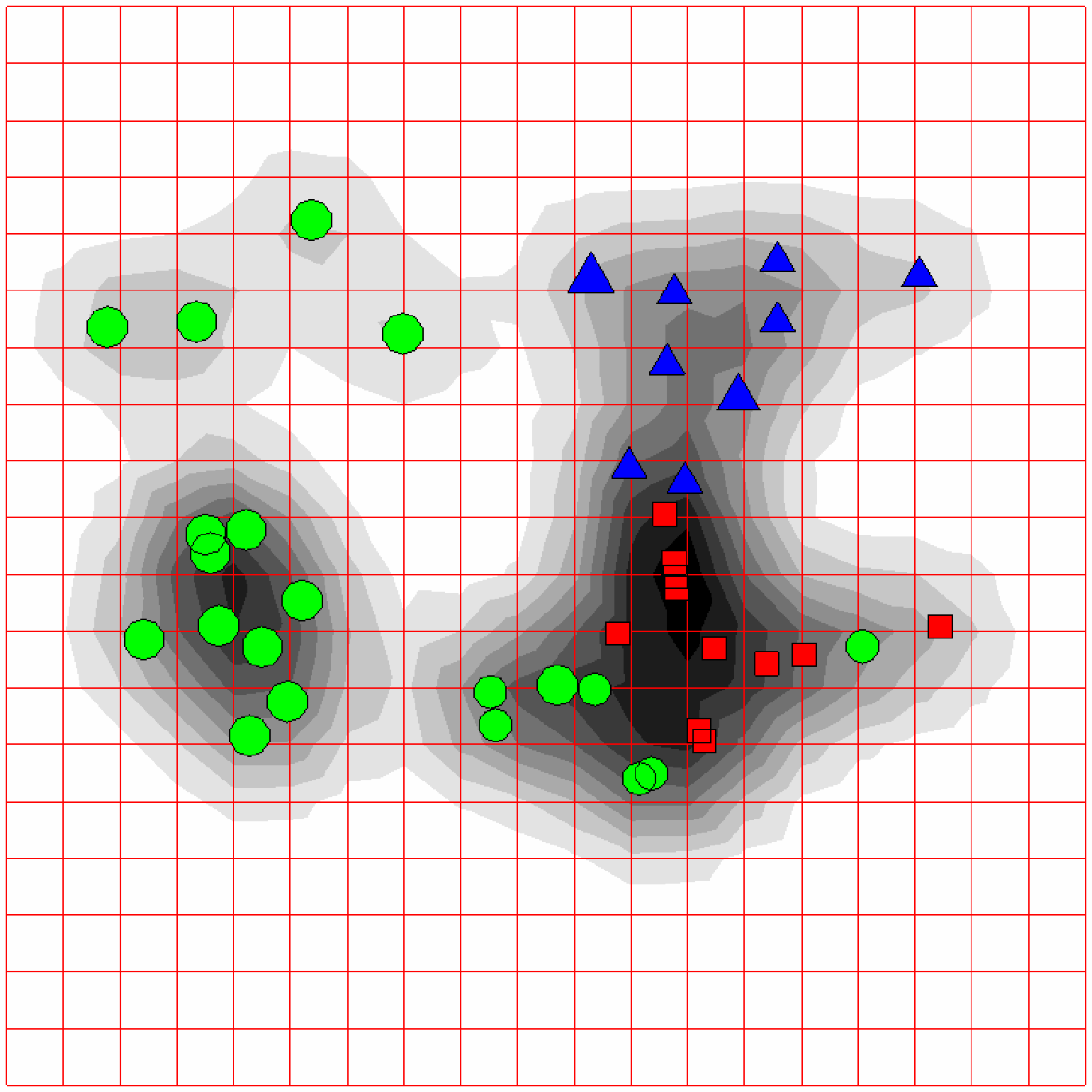}\\
\textbf{b}) ELMap2D \hspace{35mm} \textbf{c}) PCA2D }
\caption{Visualization of Dataset II (bladder cancer dataset)
using elastic maps. {\it Ab initio} classifications are shown
using points size (GRADE) and color and shape (TYPE): \textbf{a})
configuration of nodes in the three-dimensional principal linear
manifold; \textbf{b}) distribution of points in the internal
manifold coordinates is shown together with an estimation of the
two-dimensional density of points; \textbf{c}) the same as b), but
for the linear manifold of the first two principal components
(PCA2D). From all three datasets, Dataset II is ``the least
dimensional'', and the configurations of points are quite similar,
although ELMap2D reveals some finer point grouping
\label{DatasetII}}
\end{figure}

\begin{figure}
\centering{ \hspace{3mm}\includegraphics[width=90mm,
height=80mm]{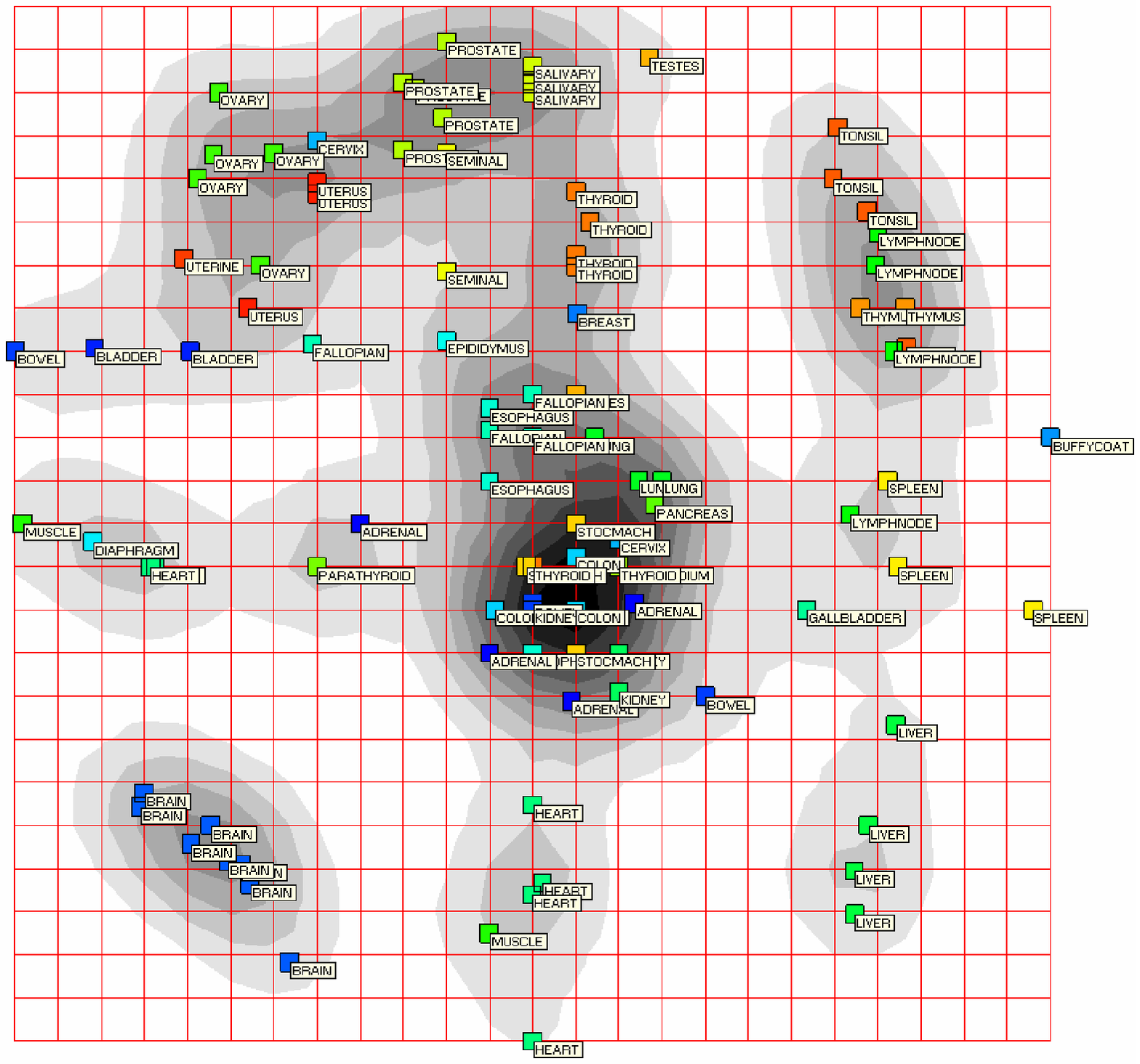}\\ \textbf{a}) ELMap2D\\
\includegraphics[width=83mm, height=80mm]{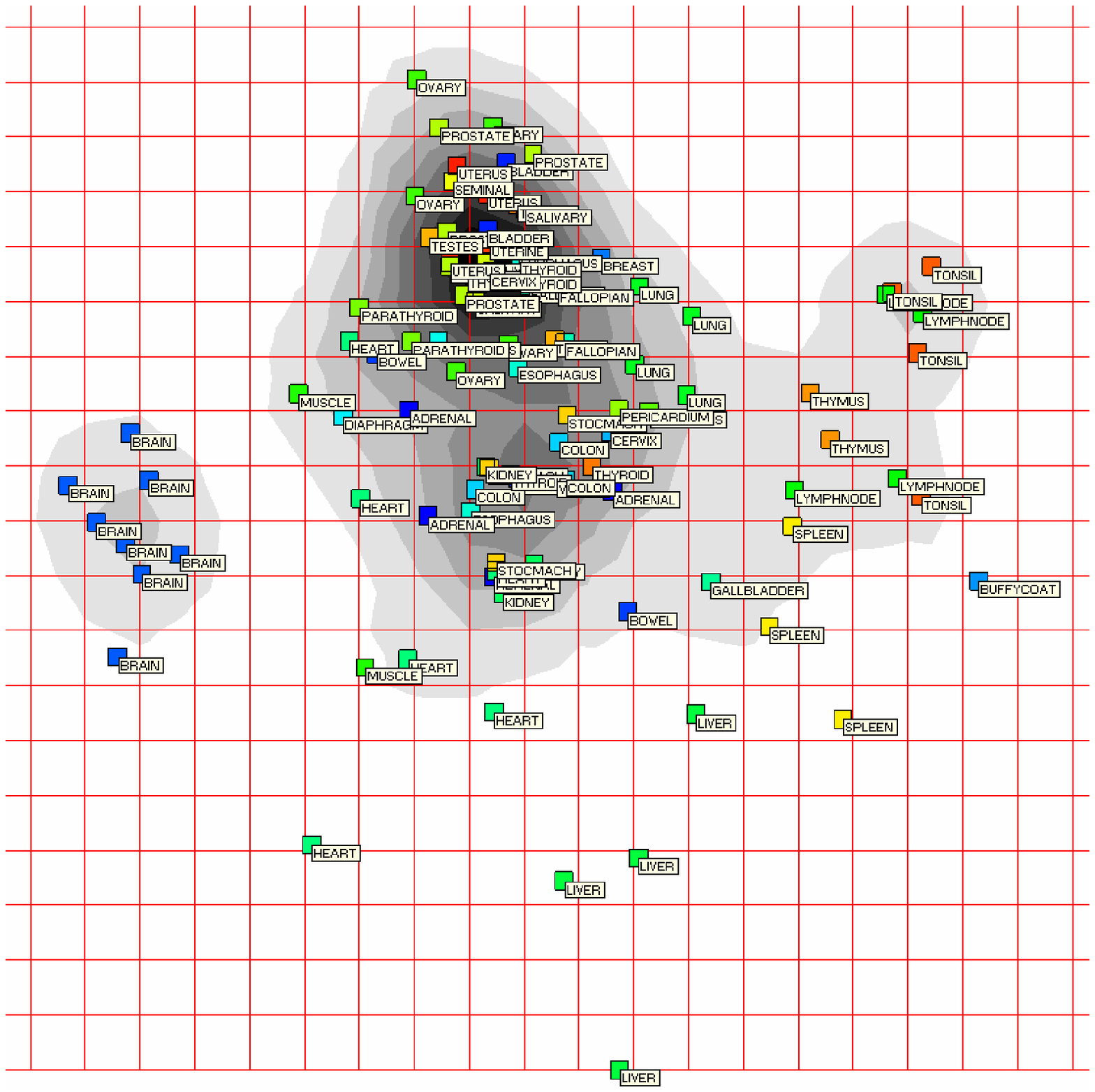}\\
b) PCA2D } \caption{Visualization of Dataset III (normal tissue
samples) using elastic maps (\textbf{a}) and two-dimensional PCA
(\textbf{b}). Tissue names are shown using point labels. The
ELMap2D image clearly gives a better ``resolved'' structure of
classes, which is confirmed by quantitative measures of ``class
compactness'' (see the text) \label{DatasetIII}}
\end{figure}

Principal manifolds for three datasets were approximated by the
elastic map algorithm implemented in Java, the package {\it
VDAOEngine } \cite{VIMIDA} with a 2D rectangular node grid using
the softening optimization strategy. The elastic nets were trained
in two epochs: first one with $\lambda_0=0.1$ and $\mu_0=250$ and
the second epoch with $\lambda_0=0.01$ and $\mu_0=30$. No
optimization for the final value of $\mu_0$ was done. The
resolution of grids was selected 15x15 for Dataset I and Dataset
III and 10x10 for Dataset II. For post-processing we used linear
extrapolation of the manifold, adding 5 node layers at each border
of the manifold. An average time for elastic net computation using
an ordinary PC was 1-5 minutes depending on the number of nodes in
the net.

For data visualization and analysis we projected data points into
their closest points on the manifold obtained by piece-wise linear
interpolation between the nodes. The results are presented on
Fig.~\ref{DatasetI}-\ref{DatasetIII}. We show also the
configuration of nodes projected into the three-dimensional
principal linear manifold. This is done only for visualization of
the fact that the elastic map is non-linear and usually better
approximates the distribution of points in comparison with the
linear principal plane. But one should not forget that the
non-linear manifold is embedded in all dimensions of the
high-dimensional space, not being limited to the space of the
first principal components.

\subsubsection{Mean-square distance error}

By their construction, principal manifolds provide better
approximation of data than the less flexible linear PCA. In the
Table \ref{MSEtable} we provide results of comparing the mean
square distance error (MSE) of the datasets' approximation by a
two-dimensional non-linear elastic map and linear manifolds of
various dimensions (1-5-dimensional and 10-dimensional)
constructed with using standard PCA. In the case of elastic maps
we calculate the distance of the projection to the closest point
of the manifold as it is described in the section
\ref{projectionsection}. The conclusion is that the
two-dimensional principal manifolds outperform even
four-dimensional linear PCA manifolds in their accuracy of data
approximation.

\begin{table}
  \caption{Mean square distance error (MSE) of data approximation by two-dimensional elastic
  maps (ELMap2D) and linear PCA manifolds of various dimensions
  (PC1-PC5,PC10). For linear principal components, the cumulative
  explained variation is also shown
  \label{MSEtable}}

  \centering{\small

\begin{tabular}{lcllllll}

\hline

 Dataset   & \;ELMap2D   & \;PC1   & \;PC2   & \;PC3   &
\;PC4
  & \;PC5   & \;PC10 \\ \hline Breast cancer MSE  & \;48.1
  & \;52.6   & \;50.7   & \;49.3   & \;48.4   & \;
47.6   & \;45.3
\\

Variation explained   & \;-   & \;7.9\%   & \;14.3\%    & \;19.0\%
& \;22.0\%   & \;24.6\%   & \;31.5\%
\\

\hline

Bladder cancer MSE   & \;40.6   & \;48.7   & \;45.4   & \;42.8
  & \;40.7   & \;39.2   & \;33.0
\\

Variation explained   & \;-   & \;21.0\%   & \;31.4\%   & \;38.9\%
& \;44.8\%   & \;48.8\%   & \;63.8\%
\\

\hline

Normal tissues MSE   & \;36.9   & \;48.8   & \;45.5   & \;42.3
  & \;40.1   & \;38.5   & \;32.4\\

Variation explained   & \;-   & \;10.7\%   & \;19.1\%   & \;26.0\%
& \;30.3\%   & \;32.2\%   & \;40.7\%
\\

\hline

\end{tabular}

  }
\end{table}

\subsubsection{Natural PCA and distance mapping}

\index{natural PCA}

Our second analysis answers the question of how successfully the
structure of distances in the initial highly multidimensional
space is reconstructed after projection into the low-dimensional
space of the internal co-ordinates of the principal manifold. A
possible indicator is the correlation coefficient between all
pair-wise distances:

\begin{equation}
r = corr(d_{ij},\hat{d}_{ij})\; ,
\end{equation}

\noindent where $d_{ij}$ and $\hat{d}_{ij}$ are the pair-wise
distances in the original high-dimensional space and the distances
in the ``reduced'' space after projection onto a low-dimensional
object (this object can be principal manifold, linear or
non-linear). However, not all of the distances $d_{ij}$ are
independent, which creates a certain bias in the estimation of the
correlation. We propose a solution for reducing this bias by
selecting a subset of ``representative'' independent distances
between the points. This is done by a simple procedure which we
call Natural PCA (NatPCA) that is described below. Let $M$ be a
finite set of points and $S\in M$ be a subset. We define a
distance from a point $i\in M$ to the set $S$ as

\begin{equation}
dist(i,S) = min \{d_{ij}, j \in S\}\; .
\end{equation}

Let us suppose that a point $i$ has the closest point $j$ in $S$. If
there are several closest points, we choose one randomly. We will
call \index{natural principal components} {\it natural principal
components} $m-1$ ($m$ is the number of points) pairs of points
$\{i,j\}\in M\times M$ that are selected by the following algorithm:

\vspace*{1mm} \frame{\hspace{4mm}
\begin{minipage}[l]{10.3cm}
\vspace{2mm}

\hspace{-3.5mm}0.~Let $S$ be an empty set.

\hspace{-3.5mm}1.~The first component is a pair of the most
distant points $\{i_m,j_m\}=\arg \sup_{ij} d_{ij}$. We put $i_m$
and $j_m$ in $S$.

\hspace{-3.5mm}2.~Among all points which are not in $S$ we select
a point $k_m$ which is the most distant to $S$:

\begin{equation}
k_m = \arg \sup_{j} \{dist(j,S)\}\; .
\end{equation}

\hspace{-3.5mm}3.~We define next the ``natural'' component as a
pair $\{k_m,p_m\}$, where $p_m\in S$ is the point in $S$, closest
to $k_m$. We add $k_m$ to $S$.

\hspace{-3.5mm}4.~We repeat steps 2-3 until all points are in $S$.
\vspace{0.1mm}
\end{minipage}
}\vspace*{1mm}

As a result we have a set NatPCA of $m-1$ pairs of points with
decreasing distances between them. These distances are independent
from each other and represent all scales in the distribution of data
(its diameter, ``second'' diameter and so on). In other words, the
first ``natural'' component is defined by the most distant points,
the second component includes the most distant point from the first
pair, and so on. We call this set {\it natural principal components}
because a) for big datasets in vector spaces the vectors defined by
the first pairs are usually located close to principal components
given that there are no outliers in the data\footnote{By its
construction, NatPCA is sensitive to presence of outliers in the
data. This problem can be solved by using pre-clustering (adaptive
coding) of data with $K$-means or $K$-medoids algorithms with
relatively big $K$ and performing NatPCA on the centroids (or
medoids). }; b) this analysis can be made in a finite metric space,
using only a distance matrix as input. NatPCA forms a tree
connecting points in the manner which is opposite to
neighbour-joining hierarchical clustering.

Here we use NatPCA to define a subset $NatPCA \in \{d_{ij}\}$ of
independent distances in the initial high-dimensional space, which
is used to measure an adequacy of distance representations in the
reduced space:

\begin{equation}
r = corr(d_{ij},\hat{d}_{ij}), \{i,j\}\in NatPCA \; .
\end{equation}

In the Table \ref{distancerep} we compare the results of calculating
$r$ using Pearson and Spearman rank correlation coefficients, $r$
calculated for the two-dimensional principal manifold and for linear
manifolds of various dimensions constructed with the use of standard
PCA. One clear conclusion is that the two-dimensional elastic maps
almost always outperform the linear manifolds of the same dimension,
being close in reproducing the distance structure to
three-dimensional linear manifolds. This is rather interesting since
the principal manifolds are not constructed to solve the problem of
distance structure approximation explicitly (unlike Multidimensional
Scaling). The results also hold in the case of using all pair-wise
distances without the application of NatPCA (data not shown),
although NatPCA selection gives less biased correlation values.

\begin{table}
  \caption{Quality of reproducing distances after dimension reduction by two-dimensional elastic
  maps (ELMap2D) and linear PCA manifolds of various dimensions (PC1-PC5,PC10)
\label{distancerep}  }
  \centering{\small

\begin{tabular}{lcllllll}

\hline

 Dataset/method & \; ELMap2D & \; PC1 & \; PC2 & \; PC3 & \;
PC4 & \; PC5 & \; PC10
\\ \hline Breast cancer/Pearson & \; 0.60 & \;  0.40 & \;  0.52 & \; 0.61 & \;
0.65 & \; 0.69 & \; 0.75 \\ Breast cancer/Spearman & \; 0.40 & \;
0.19 & \; 0.32 & \; 0.36 & \; 0.42 & \; 0.49 & \; 0.56 \\

\hline

Bladder cancer/Pearson & \; 0.87 & \;  0.82 & \;  0.84 & \; 0.88 &
\; 0.89 & \; 0.91 & \; 0.96 \\ Bladder cancer/Spearman & \; 0.68 &
\; 0.57 & \; 0.60 & \; 0.70 & \; 0.70 & \; 0.75 & \; 0.90 \\

\hline

Normal tissues/Pearson & \; 0.80 & \;  0.68 & \;  0.78 & \; 0.82 &
\; 0.86 & \; 0.87 & \; 0.95
\\ Normal tissues/Spearman & \; 0.68 & \;  0.56 & \;  0.69 & \; 0.79 & \; 0.84 & \;
0.86 & \; 0.94 \\

\hline

\end{tabular}

  }
\end{table}

\subsubsection{Local neighbourhood preservation}

\begin{figure}
\centering{ \includegraphics[width=53mm, height=43mm]{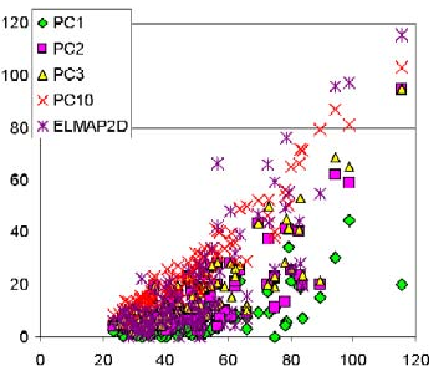}
\includegraphics[width=53mm, height=43mm]{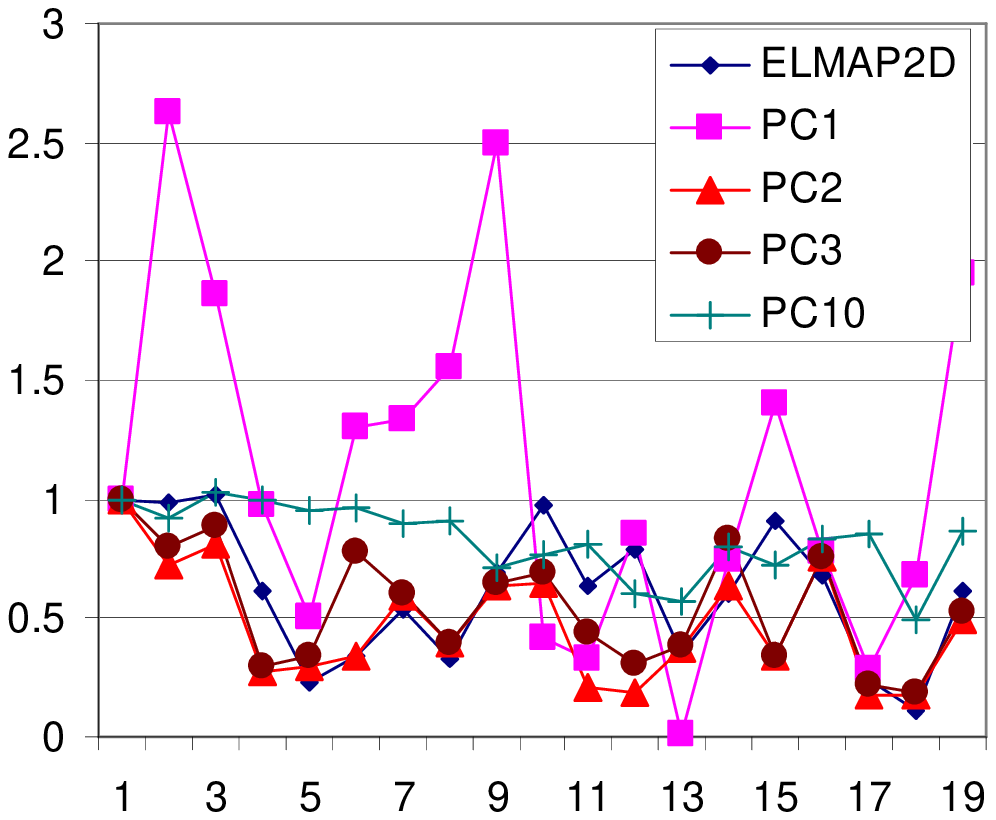}\\
\textbf{a}) \hspace{55mm} \textbf{b}) } \caption{ \textbf{a})
Sheppard plot ($\hat{d}_{ij}$ vs $d_{ij} $) for a subset of
pair-wise distances selected by NatPCA, calculated for the normal
tissue dataset where $\hat{d}_{ij}$ are calculated for dimension
reduction by elastic maps (ELMap2D), and linear principal
components (PC1-PC5, PC10); distances $\hat{d}_{ij}$ calculated in
the internal coordinates of the elastic map (ELMap2D) were scaled
such that $\hat{d}_{ij}^{ELMap2D}/d_{ij}=1$ for the first
(largest) distance; $\hat{d}_{ij}$ for linear manifolds were taken
without re-scaling; \textbf{b}) $\hat{d}_{ij}/d_{ij}$ ratios for
the first 20 $NatPCA$ components; the ratios are scaled such that
$\hat{d}_{ij}/d_{ij}=1$ for the first (largest) distance
\label{NPCABreast}}
\end{figure}

Dimension reduction using projections onto low-dimensional linear
manifolds leads to distance contraction, i.e. all the pair-wise
distances become smaller than they were in the initial
multidimensional space. The structure of big distances is reproduced
relatively well since linear principal components are aligned along
the directions of the biggest variation. This is demonstrated on
Fig.~\ref{NPCABreast}, where scatters of $\{d_{ij},\hat{d}_{ij}\}$
(so-called Sheppard plots) and scaled ratios $\hat{d}_{ij}/d_{ij}$
are shown for distances from NatPCA set. One can expect that
$k$-dimensional PCA is able to reproduce correctly the first $k$
``natural'' principal components, and more if the dimension of the
data is less than $k$.

However, preservation of the local structure of small distances is
generally not guaranteed, hence it is interesting to measure it. We
used a simple test which consists of calculating a set of $k$
neighbors in the ``reduced'' space  for every point, and counting
how many of them are also point neighbors in the initial
high-dimensional space. The measure of local distance structure
preservation is the average ratio of the intersection size of these
two sets ($k$ neighbors in the ``reduced'' and the initial space)
over $k$. The unit value would correspond to perfect neighborhood
preservation, while a value close to zero corresponds to strong
distortions or a situation where distant regions of dataspace are
projected in the same region of the low-dimensional manifold. The
results are provided in Table \ref{LocNeighTable}. The elastic maps
perform significantly better than the two-dimensional linear
principal manifold in two cases and as well as two-dimensional PCA
for Dataset II. To evaluate the significance of this improvement, we
made a random test to calculate the local neighborhood preservation
measure obtained by a pure random selection of points.

\begin{table}
  \caption{Local neighbourhood preservation (defined as an average relative number
  of neighbours, ``preserved'' after dimension reduction) for two-dimensional (2D) elastic
  maps (ELMap) and linear PCA manifolds of dimensions 1-5 and 10
  (PC1-PC5, PC10). The last column (RANDOM) corresponds to the value
  calculated for a random choice of neighbours (10000 permutations are made)
  which is shown together with its standard deviation
  \label{LocNeighTable}}

  \centering{\small

\begin{tabular}{lclllllll}

\hline

Dataset & \, ELMap & \, PC1 & \, PC2 & \, PC3 & \, PC4 & \, PC5 &
\, PC10 &  RANDOM
\\ \hline Breast cancer (k=10) & \, 0.26 & \,  0.13 & \,  0.20 & \, 0.28 & \,
0.31 & \, 0.38 & \, 0.47 &  0.04 $\pm$ 0.06
\\

\hline

Bladder cancer (k=5) & \, 0.53 & \,  0.34 & \,  0.53 & \, 0.61 &
\, 0.64 & \, 0.70 & \, 0.80 & 0.12 $\pm$ 0.14
\\

\hline

Normal tissues (k=5) & \, 0.49 & \,  0.23 & \,  0.33 & \, 0.43 &
\, 0.50 & \, 0.54 & \, 0.69 &  0.05 $\pm$ 0.09
\\
\hline

\end{tabular}
  }
\end{table}

\subsubsection{Class compactness}

\index{compactness}

Now let us suppose that the points are marked by labels, thus
defining classes of points. In the process of dimension reduction,
various distortions of the data space are possible. Distant points
can be closely projected, distorting the picture of inter and
intra-class distances existing in the multidimensional space. In
this section we analyze how principal manifolds preserve class
distance relations in comparison with linear principal manifolds of
various dimensions using a simple test for ``class compactness''.

For a class $C$, let us define ``class compactness'' as an average
of a proportion of the points of class $C$ among $k$ neighbors of
the point. This average is calculated only over the points from
the class $C$.

In Table \ref{ClassTable}, we provide values of ``class
compactness'' for all datasets with different {\it ab initio}
classes defined. One can see that there are many examples when
two-dimensional principal manifolds perform as well as four- or
five-dimensional linear manifolds in putting together the points
from the same class. It works particularly well for the collection
of normal tissues. There are cases when neither linear nor
non-linear low-dimensional manifolds could put together points of
the same class and there are a few examples when linear manifolds
perform better. In the latter cases (Breast cancer's A, B, lumA,
lumB and ``unclassified'' classes, bladder cancer T1 class),
almost all class compactness values are close to the estimated
random values which means that these classes have big intra-class
dispersions or are poorly separated from the others. In this case
the value of class compactness becomes unstable (look, for
example, at the classes A and B of the breast cancer dataset) and
depends on random factors which can not be taken into account in
this framework.

The closer class compactness is to unity, the easier one can
construct a decision function separating this class from the
others. However, in the high-dimensional space, due to many
degrees of freedom, the ``class compactness'' might be compromised
and become better after appropriate dimension reduction. In Table
\ref{ClassTable} one can find examples when dimension reduction
gives better class compactness in comparison with that calculated
in the initial space (breast cancer basal subtype, bladder cancer
Grade 2 and T1, T2+ classes). It means that sample classifiers can
be regularized by dimension reduction using PCA-like methods.

There are several particular cases (breast cancer basal subtype,
bladder cancer T2+, Grade 2 subtypes) when non-linear manifolds
give better class compactness than both the multidimensional space
and linear principal manifolds of the same dimension. In these
cases we can conclude that the dataset in the regions of these
classes is naturally ``curved'' (look, for example, at
Fig.~\ref{DatasetI}) and the application of non-linear techniques
for classification regularization is an appropriate solution.

We can conclude that non-linear principal manifolds provide
systematically better or equal resolution of class separation in
comparison with linear manifolds of the same dimension. They
perform particularly well when there are many small and relatively
compact heterogeneous classes (as in the case of normal tissue
collection).

\begin{table}
  \caption{``Class compactness'' before (ND) and after dimension
  reduction with use of two-dimensional (2D) elastic maps (ElMap) and
  linear principal manifolds for 1-5 and 19 principal components
  (PC1-PC5,PC10). The column ``Random''
  shows a random test when $k$ points are selected by chance
  (10000 permutations have been made). The number
  of samples in the class is shown in parentheses (only classes with
  $>$ 2 samples are shown); ``unclas." stands for unclassified samples
\label{ClassTable}  }
  \centering{ \small
\begin{tabular}{llcllllllr}

\hline

Class & ND & ElMap & PC1\, & PC2\, & PC3\, & PC4\, & PC5\, & PC10
& Random\,
\\ \hline
\multicolumn{10}{c}{Breast cancer (GROUP), k=5}\\

A (193 samples)&0.73& 0.67&
0.71&0.67&0.68&0.69&0.70&0.70&0.67$\pm$0.27
\\ B (93 samples)&0.31& 0.30&
0.33&0.29&0.35&0.37&0.38&0.31&0.33$\pm$0.21 \\ \hline
\multicolumn{10}{c}{Breast cancer (TYPE), k=3}\\

lumA (95)&0.61&0.60&0.64&0.65&0.67&0.72&0.71&0.74&0.33$\pm$0.27
\\ basal (55)&0.88&0.93&0.83&0.86&0.88&0.90&0.88&0.89&0.19$\pm$0.17\\ unclas.
(42)&0.28&0.20&0.27&0.25&0.27&0.27&0.30&0.29&0.15$\pm$0.16\\
normal
(35)&0.68&0.28&0.14&0.19&0.31&0.41&0.41&0.55&0.12$\pm$0.19\\ errb2
(34)&0.62&0.36&0.24&0.34&0.32&0.32&0.43&0.59&0.12$\pm$0.19\\ lumB
(25)&0.21&0.08&0.20&0.20&0.21&0.25&0.23&0.36&0.09$\pm$0.17\\
\hline \multicolumn{10}{c}{Bladder cancer (TYPE), k=3}\\

Ta (20)&0.85&0.83&0.80&0.85&0.85&0.85&0.85&0.85&0.50$\pm$0.29\\ T1
(11)&0.58&0.67&0.45&0.69&0.67&0.70&0.67&0.63&0.27$\pm$0.26\\ T2+
(9)&0.85&0.89&0.11&0.85&0.81&0.89&0.78&0.81&0.22$\pm$0.24\\ \hline
\multicolumn{10}{c}{Bladder cancer (GRADE), k=3}\\

Grade 3 (32)\,
&0.94&0.92&0.86&0.83&0.90&0.95&0.94&0.94&0.8$\pm$0.23\\ Grade 2
(6)&0.5&0.61&0.22&0.39&0.67&0.67&0.67&0.61&0.15$\pm$0.23\\

\hline \multicolumn{10}{c}{Normal tissues (TISSUE), k=2}\\

BRAIN (8)&1.0&1.0&1.0&0.94&1.0&1.0&1.0&1.0&0.08$\pm$0.19\\

HEART (6)&0.92&0.50&0.17&0.25&0.25&0.25&0.33&0.58&0.06$\pm$0.16\\

THYROID
(6)&0.92&0.67&0.17&0.17&0.25&0.17&0.67&0.75&0.06$\pm$0.17\\

PROSTATE (5)&1.0&0.8&0.0&0.0&0.1&0.4&0.5&1.0&0.05$\pm$0.16\\

OVARY (5) & 0.7 & 0.6 & 0.2 & 0.1 & 0.3 & 0.4 & 0.5 & 0.7 &
0.05$\pm$0.15\\

LUNG (4)&0.88&0.40&0.5&0.63&0.75&0.63&0.63&0.75&0.04$\pm$0.13\\

ADRENAL (4)&0.625&0.0&0.0&0.0&0.0&0.25&0.5&0.625&0.04$\pm$0.14\\

LIVER (4)&0.88&0.88&0.13&0.75&0.75&0.88&0.88&0.88&0.04$\pm$0.14\\

SALIVARY (4)&1.0&1.0&0.25&0.25&0.50&0.88&1.0&1.0&0.04$\pm$0.14\\

TONSIL (4)&0.88&0.63&0.0&0.13&0.34&0.75&0.75&0.88&0.04$\pm$0.14\\

LYMPHNODE (4) \;
&0.75&0.25&0.25&0.25&0.0&0.25&0.25&0.25&0.04$\pm$0.14\\

FALLOPIAN
(4)&0.88&0.50&0.0&0.13&0.25&0.25&0.25&0.75&0.04$\pm$0.14\\

SPLEEN (3)&0.67&0.50&0.50&0.33&0.33&0.89&0.67&0.83&0.03$\pm$0.12\\

KIDNEY (3)&1.0&0.33&0.0&0.0&0.33&0.33&0.33&0.67&0.03$\pm$0.12\\

UTERUS (3)&0.83&0.33&0.0&0.33&0.33&0.33&0.33&0.83&0.03$\pm$0.12\\

BOWEL (3)&0.17&0.0&0.0&0.0&0.0&0.0&0.0&0.0&0.03$\pm$0.12\\

ESOPHAGUS (3)&0.67&0.0&0.0&0.0&0.0&0.0&0.0&0.67&0.03$\pm$0.12\\

COLON (3) & 0.33 & 0.0 & 0.17 & 0.0 & 0.0 & 0.17 & 0.17 & 0.17 &
0.03$\pm$0.12 \\

STOCMACH (3)&0.5&0.0&0.0&0.0&0.0&0.0&0.0&0.0&0.03$\pm$0.12\\

\hline

\end{tabular}
  }
\end{table}

\section{Discussion}

Principal Components Analysis already celebrated its 100th
anniversary \cite{4Pearson1901}. Nowadays linear principal manifolds
are widely used for dimension reduction, noise filtering and data
visualization. Recently, methods for constructing non-linear
principal manifolds were proposed, including our {\it elastic maps}
approach which is based on a physical analogy with elastic
membranes. We have developed a general geometric framework for
constructing ``principal objects'' of various dimensions and
topologies with the simplest quadratic form of the smoothness
penalty, which allows very effective parallel implementations.

Following the metaphor of elasticity (Fig.~\ref{4Fig3}), we
introduced two smoothness penalty terms, which are quadratic at the
node optimization step. As one can see from
(\ref{ElasticFunctional}) they are similar to the sum of squared
grid approximations of the first and second derivatives \footnote{
The differences should be divided by node-to-node distances in order
to be true derivative approximations, but in this case the quadratic
structure of the term would be violated. We suppose that the grid is
regular with almost equal node-to-node distances, then the
dependence of coefficients \textit{$\lambda $}$_{i}$, \textit{$\mu
$}$_{j}  $ on the total number of nodes contains this factor.} . The
$U^\phi_{E}$ term penalizes the total length (or area, volume) of
the principal manifold and, indirectly, makes the grid regular by
penalizing non-equidistant distribution of nodes along the grid. The
$U^\phi_{R}$ term is a smoothing factor, it penalizes the
nonlinearity of the ribs after their embedding into the data space.

Such quadratic penalty allows using standard minimization of
quadratic functionals (i.e., solving a system of linear algebraic
equations with a sparse matrix), which is considerably more
computationally effective than gradient optimization of more
complicated cost functions, like the one introduced by K\'{e}gl.
Moreover, the choice of (\ref{ElasticFunctional}) is the simplest
smoothness penalty, which is universally applicable.

Minimization of a positive definite quadratic functional can be
provided by the sequential one-dimensional minimization for every
space coordinate (cyclic). If for a set of coordinates $\{x_i\}_{i
\in J}$ terms $x_i x_j$ ($i,j \in J$, $i \neq j$) do not present in
the functional, then for these coordinates the functional can be
minimized independently. The quadratic functional we formulate has a
sparse structure, it gives us the possibility to use parallel
minimization that is expected to be particularly effective in the
case of multidimensional data. The application of high-throughput
techniques in molecular biology such as DNA microarrays provides
such data in large quantities. In the case when $n\gg m$ (the space
dimension is much bigger than the number of samples) many existing
algorithms for computation of principal manifolds (as GTM
\cite{4Bishop1998}) change their performance. In particular,
computing the closest nodes with the formula (\ref{taxondef}) is no
longer the limiting step with most computations being required for
finding the new node positions at each iteration.

Our approach is characterized by a universal and flexible way to
describe the connection of nodes in the grid. The same algorithm,
given an initial definition of the grid, provides construction of
principal manifolds with various dimensions and topologies. It is
implemented together with many supporting algorithms
(interpolation/extrapolation formulas, adaptation strategies, and
so on) in several programming languages
\cite{Elmap,VidaExpert,VIMIDA}.

One important application of principal manifolds is dimension
reduction and data visualization. In this field they compete with
multidimensional scaling methods and recently introduced
algorithms of dimension reduction, such as the locally linear
embedding (LLE) \cite{Roweis00} and ISOMAP \cite{Tenenbaum00}
algorithms. The difference between these two approaches is that
the later ones seek new point coordinates directly and do not use
any intermediate geometrical objects. This has several advantages,
in particular that a) there is a unique solution for the problem
(the methods are not iterative in their nature, there is no
problem of grid initialization) and b) there is no problem of
choosing a good way to project points onto a non-linear manifold.
Another advantage is that the methods are not limited by several
first dimensions in dimension reduction (it is difficult in
practice to manipulate non-linear manifolds of dimension more than
three). However, when needed we can overcome this limitation by
using an iterative calculation of low-dimensional objects.

A principal manifold can serve as a non-linear low-dimensional
screen to project data. The explicit introduction of such an
object gives additional benefits to users. First, the manifold
approximates the data and can be used itself, without applying
projection, to visualize different functions defined in data space
(for example, density estimation). Also the manifold as a mediator
object, ``fixing'' the structure of a learning dataset, can be
used in visualization of data points that were not used in the
learning process, for example, for visualization of dataflow ``on
the fly''. Constructing manifolds does not use a point-to-point
distance matrix, this is particularly useful for datasets with
large $n$ when there are considerable memory requirements for
storing all pair-wise distances. Also using principal manifolds is
expected to be less susceptible  to additive noise than the
methods based on the local properties of point-to-point distances.
To conclude this short comparison, LLE and ISOMAP methods are more
suitable if the low-dimensional structure in multidimensional data
space is complicated but is expected, and if the data points are
situated rather tightly on it. Principal manifolds are more
applicable for the visualization of real-life noisy observations,
appearing in economics, biology, medicine and other sciences, and
for constructing data screens showing not only the data but
different related functions defined in data space.

In this paper we demonstrated advantages of using non-linear
principal manifolds in visualization of DNA microarrays, data
massively generated in biological laboratories. The non-linear
approach, compared to linear PCA which is routinely used for
analysis of this kind of data, gives better representation of
distance structure and class information.

\end{document}